\documentclass[aps,prd,twocolumn,notitlepage,
superscriptaddress,
nofootinbib,floatfix]{revtex4-2}
\usepackage[utf8]{inputenc} %Support for additional characters in bibliography
\usepackage{amsfonts,amssymb, mathrsfs, amsmath, esint,bm,enumitem}
\usepackage{subcaption}
\usepackage{graphicx}
\allowdisplaybreaks

\usepackage{subcaption}
\usepackage{ragged2e}
\DeclareCaptionJustification{justified}{\justifying}
\makeatletter    
\renewcommand\onecolumngrid{% <<<<<<
	\do@columngrid{one}{\@ne}%
	\def\set@footnotewidth{\onecolumngrid}% <<<<<<<<<<<<<<<<
	\def\footnoterule{\kern-6pt\hrule width 1.5in\kern6pt}%
}
\renewcommand\twocolumngrid{% <<<<<<
	\def\footnoterule{% restore rule
		\dimen@\skip\footins\divide\dimen@\thr@@
		\kern-\dimen@\hrule width.5in\kern\dimen@}
	\do@columngrid{mlt}{\tw@}
}%
\makeatother 

\usepackage{diagbox}
\usepackage{latexsym}
\usepackage{graphicx}
\usepackage[dvipsnames]{xcolor}
\usepackage{booktabs,multirow}
\usepackage{titlesec} % to change style of \paragraph{} header
\usepackage{physics}
\usepackage[normalem]{ulem}
\graphicspath{{./Figures/}}

\usepackage{hyperref}
\hypersetup{colorlinks, citecolor=ForestGreen, linkcolor=BrickRed, urlcolor=RedViolet}

\newcommand{\ann}{\text{ann}}
\newcommand{\DV}{{\Delta V}}

\begin{document}

	\title{Collapsing Domain Wall Networks: 
		\\ Impact on Pulsar Timing Arrays and Primordial Black Holes }

	\author{Ricardo Z. Ferreira} 
	\email{rzferreira@uc.pt}
	\affiliation{CFisUC, Department of Physics, University of Coimbra, P-3004 - 516 Coimbra, Portugal}
	\author{Alessio Notari} 
	\email{notari@fqa.ub.edu}
	\affiliation{Departament de F\'isica Qu\`antica i Astrofis\'ica \& Institut de Ci\`encies del Cosmos (ICCUB), Universitat de Barcelona, Mart\'i i Franqu\`es 1, 08028 Barcelona, Spain. 
		\looseness=-1}
	\affiliation{Galileo Galilei Institute for theoretical physics,
		Centro Nazionale INFN di Studi Avanzati Largo Enrico Fermi 2, I-50125, Firenze, Italy
		\looseness=-1}
	\author{Oriol Pujolàs} 
	\email{pujolas@ifae.es}
	\affiliation{Institut de F\'isica d’Altes Energies (IFAE) and The Barcelona Institute of Science and Technology (BIST), Campus UAB, 08193 Bellaterra (Barcelona), Spain}
	\author{Fabrizio Rompineve} 
	\email{frompineve@ifae.es}
	\affiliation{Departament de F\'isica, Universitat Aut\`onoma de Barcelona, 08193 Bellaterra, Barcelona, Spain
		\looseness=-1}
	\affiliation{Institut de F\'isica d’Altes Energies (IFAE) and The Barcelona Institute of Science and Technology (BIST), Campus UAB, 08193 Bellaterra (Barcelona), Spain}
	\affiliation{CERN, Theoretical Physics Department,
		Esplanade des Particules 1, Geneva 1211, Switzerland}

\begin{abstract}
\noindent

Unstable domain wall (DW) networks in the early universe are cosmologically viable and can emit a large amount of gravitational waves (GW) before annihilating. As such, they provide an interpretation for the recent signal reported by Pulsar Timing Array (PTA) collaborations. A related important question is whether such a scenario also leads to significant production of Primordial Black Holes (PBH). 
We investigate both GW and PBH production using 3D numerical simulations in an expanding background, with box sizes up to $N=3240$,
including the annihilation phase. We find that: 
i)  the network decays exponentially, i.e. the false vacuum volume drops as $\sim \exp(-\eta^3)$, with $\eta$ the conformal time;
ii) the GW spectrum is larger than traditional estimates by more than one order of magnitude, due to a delay between DW annihilation and the sourcing of GWs. 
We then present a novel semi-analytical method to estimate the PBH abundances: rare  false vacuum pockets of super-Hubble size collapse to PBHs if their energy density becomes comparable to the background  when they cross the Hubble scale. Smaller (but more abundant) pockets will instead collapse only if they are close to spherical. This introduces very large uncertainties in the final PBH abundance. The first phenomenological implication is that the DW interpretation of the PTA signal is compatible with observational constraints on PBHs, within the uncertainties.
Second, in a different parameter region, the dark matter can be entirely in the form of asteroid-mass PBHs from the  DW collapse. Remarkably, this would also lead to a GW background in the observable range of LIGO-Virgo-KAGRA and future interferometers, such as LISA and Einstein Telescope.

\end{abstract}

\hfill CERN-TH-2024-020

\maketitle
\vspace{1em}\noindent

%%%%%%%%%%%%%%%%%%%%%%%%%%%%%%%%%%%%%%%%%%%%%%%%%%%%%%%%%%%%%%%%%%%%%%%%
%\newpage
\section{Introduction}
\label{sec:introduction}

Physical models that feature the formation of cosmic domain wall (DW) networks have typically been seen as  problematic due to the so-called \textit{domain wall problem}, the fact that the network tends to dominate the energy density of the universe 
 \cite{Zeldovich:1974uw}. However, if domain walls are biased and annihilate,  
 this \textit{problem} turns into a  virtue,   
  as the network naturally tends to be an abundant component in the universe before its collapse, and is thus easier to probe.

Gravitational waves (GWs) are one of the potential signatures~\cite{Preskill:1991kd,Chang:1998tb, Gleiser:1998na}. The spectrum is stochastic and analyses with LIGO-Virgo O3 data already place constraints on the parameters of the network \cite{KAGRA:2021kbb}. 
Recently, the evidence for nano-Hz GWs at Pulsar Timing Arrays (PTAs)~\cite{NG15-SGWB, EPTA2-SGWB, PPTA3-SGWB, CPTA-SGWB} brought renewed interest in this possibility: DW networks that annihilate around the QCD phase transition provide a good explanation of the signal and outperform several other models~\cite{Ferreira:2022zzo, NANOGrav:2023hvm}. In case of detection of GWs, it is also crucial to find additional signatures, that can help in selecting DWs over other early Universe sources. Dark radiation and collider signals may be some of such signatures~\cite{Ferreira:2022zzo}, as well as the production of Primordial Black Holes (PBHs) from the collapsing network~\cite{Vachaspati:2017hjw, Ferrer:2018uiu} (see also~\cite{Ge:2019ihf, Gelmini:2022nim, Gelmini:2023ngs, Gouttenoire:2023gbn}), although the resulting PBH abundance is subject to large uncertainties.

Overall, these aspects  motivate a detailed investigation of the evolution of a DW network during its collapse phase (see~\cite{Hindmarsh:1996xv,Larsson:1996sp,Correia:2014kqa, Correia:2018tty, Pujolas:2022qvs, Chang:2023rll} for previous work) 
and of its gravitational relics, i.e.
 GWs (see~\cite{Hiramatsu:2010yz, Kawasaki:2011vv, Hiramatsu:2013qaa, ZambujalFerreira:2021cte, Kitajima:2023cek,Ferreira:2023jbu}) and PBHs, which we aim to
perform in this work.
We consider the simple case of DWs with a double-well potential, with an energy per unit area (tension) $\sigma$,  and where the potential is slightly tilted by a  term of size $\Delta V$, i.e. a~{\it bias} such that one of the $\mathbb{Z}_2$-symmetric minima becomes a false vacuum~\cite{Sikivie:1982qv} (see e.g.~\cite{Vilenkin:1981zs, Coulson:1995nv, Babichev:2021uvl, Gonzalez:2022mcx} for other mechanisms to have viable long-lived walls). We simulate the corresponding DW network throughout the formation, scaling and annihilation regimes, using field theory numerical simulations in an expanding radiation-dominated background, with box size up to $3240^3$, while computing the GWs radiated throughout the evolution. 

Until recently, most analyses assumed that GWs in this scenario are radiated until the pressure on the walls caused by the bias overcomes the self-acceleration due to the wall's tension, i.e. when $\Delta V= \sigma H$, where $H$ is the Hubble rate.\footnote{A refined numerical estimate for dark matter production from string-wall networks was obtained in~\cite{Kawasaki:2014sqa}.}  Before this epoch, the network is in the so-called scaling regime, with most of its energy density in a fixed $O(1)$ number of walls per Hubble patch. However, the collapsing network consists of large DWs of various shapes, which contain False Vacuum (FV) pockets, that shrink to small sizes after the scaling epoch. These last stages of evolution can certainly source GWs in addition to the ones from the so-called scaling epoch, and has been found in recent work~\cite{Kitajima:2023cek}. 
An order-one change in the estimate of the time scale for GW production can lead to an order-of-magnitude enhancement of the final GW signal. Our simulations improve on the determination of such time-scale, while also providing new insights into the properties of the network at the onset of annihilation and highlighting the role of the kinetic energy of the scalar field in the production of gravitational waves in the final stages of the network annihilation.

Our numerical results will also allow us to establish the time evolution of the decaying network, in particular of its FV pockets.
Indeed, the collapse of the network takes some time: the abundance of Hubble-sized FV pockets at some point drops very quickly, 
but a small fraction of rare super-Hubble sized pockets survive for a longer time as they must cross the Hubble radius to annihilate.
Their abundance dramatically decreases in time, but their likelihood to collapse into a BH grows instead simply because the Schwarzschild radius associated with the FV pocket grows faster in time than the Hubble length. This may result in a tiny population of BHs at formation, but potentially large at present if the network collapses in the early universe (this formation mechanism is similar to the one in~\cite{Garriga:2015fdk,Deng:2016vzb} for isolated DWs, see also~\cite{Rubin:2000dq}, with the crucial difference that for a network the collapse is in general far from spherical). 

We provide an analytical understanding of this picture, which complements our numerical findings to provide an important step forward in the estimate of the PBH abundance. Nonetheless, large uncertainties in the final PBH abundance persist, due to an exponential sensitivity to parameters and the departure from spherical collapse, which at this point is still difficult to quantify.

With these new estimates of both GW and PBH relics, we then analyze the phenomenological consequences of a generic DW network that annihilates at different epochs in the early Universe.  First, we assess the viability of the DW interpretation of the PTA signal in light of PBH production. Second, we discuss PBHs from DWs as a candidate for the observed dark matter, with a possible correlated GW signal at interferometers such as LIGO-Virgo-KAGRA (LVK) \cite{KAGRA:2013rdx}, LISA \cite{2017arXiv170200786A}, Einstein Telescope (ET)~\cite{Punturo:2010zz} and Cosmic Explorer (CE)~\cite{Reitze:2019iox}.

The paper is organized as follows. We summarize the evolution of annihilating DW networks in Sec.~\ref{sec:networks}, highlighting the important time scales in the problem. We present the results from our numerical simulations in Sec.~\ref{sec:numerics}. We present an analytical model to account for the late FV pockets in Sec.~\ref{sec:analytics}. We conclude in Sec.~\ref{sec:pheno} with a discussion on the phenomenological impact.

\begin{figure*}\centering
	\includegraphics[width=.85\textwidth]
	{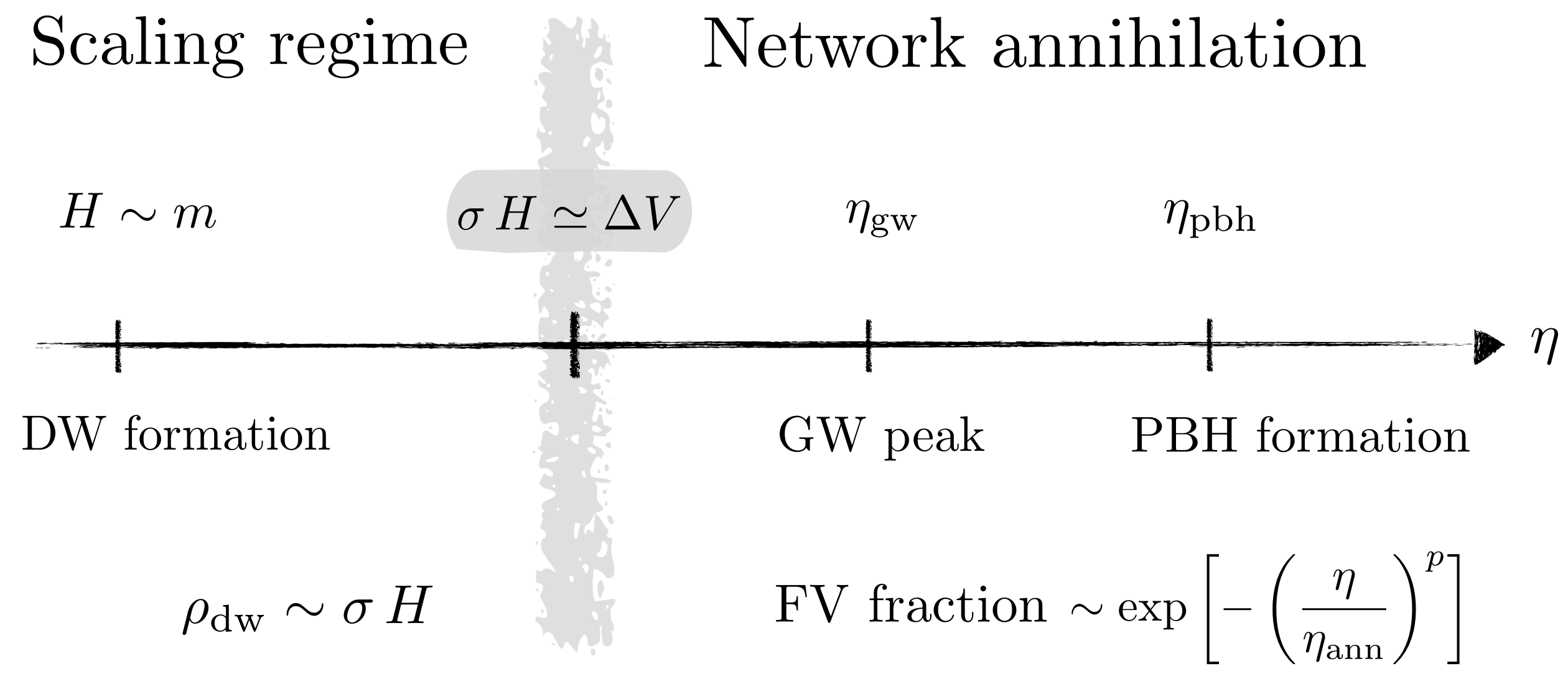}
	\caption{Timeline of DW network evolution. After spending some time in the scaling regime, the bias (vacuum energy difference $\Delta V$) becomes effective at (conformal) time $\eta_{\Delta\text{V}}$. GW and PBH production occur somewhat later, at $\eta_\text{gw}$ and $\eta_{\text{pbh}}$ respectively. The decay of the False Vacuum volume fraction, ${\cal F}_\text{fv}$, is parameterized by a decay time $\eta_\text{ann}$, slightly larger than $\eta_{\Delta\text{V}}$, and an exponent $p$. Both our numerical simulations and analytical model point to $p=3$.}
	\label{fig:DWtimeline}
\end{figure*}

%%%%%%%%%%%%%%%%%%%%%%%%%%%%%%%%%%%%%%%%%%%%%%%%%%%%%%%%%%%%%%%%%%%%%%%

%%%%%%%%%%%%%%%%%%%%%%%%%%%%%%%%%%%%%%%%%%%%%%%%%%%%%%%%%%%%%%%%%%%%%%%%
\section{Domain Wall Networks:  Scaling and Annihilation regimes}
\label{sec:networks}

In this work, we focus on a simple model exhibiting DWs: a real scalar $\phi$ with a $\mathbb{Z}_2$ symmetry $\phi\to-\phi$ and with a double well potential   
\begin{equation}\label{V}
V(\phi) =\frac{\lambda}{4}\left(\phi^2 -v^2\right)^2~,   \end{equation}
keeping in mind that several aspects of our discussion may also apply to other models (e.g.~with more minima or from axion potentials).
In this model the DW tension - the energy per unit area of the walls - is $\sigma = \frac{\sqrt{2\,\lambda}}{3} v^3$ and the scalar mass squared in the minima is $m^2 = 2 \lambda v^2$. In addition, we will assume a small bias term in the potential that breaks such a symmetry, of size $\Delta V$, that will lead to annihilation of the walls.

We assume that a network of walls is formed by a phase transition  in the early universe during radiation domination, i.e.~we start with zero initial field plus small random fluctuations and the walls are formed via the so-called Kibble mechanism~\cite{Zeldovich:1974uw, Kibble:1976sj}. The subsequent evolution follows a sequence of events, schematically represented in Fig.~\ref{fig:DWtimeline}, and explained in detail below: (i) the network reaches the so-called scaling regime, (ii) the network starts annihilating when the bias term becomes relevant, (iii) a peak of gravitational waves is produced, (iv) some very rare domain walls that survive for a longer time may collapse and give rise to PBHs.

\subsection{Scaling regime}
Soon after formation, the DW network reaches a self-similar `scaling' regime characterized by, on average, about one Hubble-sized DW per Hubble volume. If we consider a generic network of total comoving area $A$ in a box of comoving volume $V$, its total energy is $\sigma A a^2$ and so its total energy density is given by
\begin{equation}
\label{eq: rhoA}
\rho_{\text{dw}} = \frac{\sigma A a^2}{V a^3}= 2 {\cal A}\,  \sigma  H \, , 
\end{equation}
where $a$ is the scale factor, $H$ is the Hubble parameter and we introduced the so-called area parameter
\begin{equation}
\mathcal{A}\equiv \frac{A}{V}\frac{1}{2 a H } \, ,
\label{eq:Area}
\end{equation}
which is a dimensionless number related to the area density of the network. During scaling ${\cal A}$ is expected to be of order unity \cite{Hiramatsu:2013qaa}.

One of the remarkable signatures of the DW network is the stochastic spectrum of GWs that it creates. 
 In the scaling regime, the spectrum $\Omega_\text{gw}(k,t)\equiv \rho_\text{c}^{-1} \, d \rho_\text{gw}/d \log k$, where $\rho_\text{gw}$ is the energy density in GWs and  $\rho_c$ is the critical background density, as a function of  the wave number $k$ and cosmic time $t$, peaks at the Hubble scale and previous studies have found that in scaling its amplitude at the peak is given by \cite{Saikawa:2017hiv}
\begin{eqnarray} \label{eq: GW abundance}
\Omega_\text{gw}^\text{(scaling)}(k_\text{peak},t) = \frac{3}{32 \pi} {\epsilon}\,  \alpha(t)^2 \, ,
\end{eqnarray}
where $\epsilon \simeq {\cal O}(1)$ is an efficiency factor extracted from the numerical simulations and $\alpha (t)\equiv \rho_{\text{dw}}/\rho_{c}$ is the fraction of the total energy density stored in the walls. The fraction $\alpha$ increases over time and thus one expects the GW spectrum to have a  peak around the annihilation time of the network.

\subsection{Annihilation Phase}

The annihilation phase occurs roughly once the Hubble rate becomes smaller than the pressure acceleration, that is, for conformal time $\eta\gtrsim \eta_{\Delta V}$, defined by 
\begin{equation}\label{eq: annihilation condition}
    H(\eta_{\Delta V}) = \Delta V /\sigma \, ,
\end{equation}
one then expects the network to start collapsing.

So far in most literature
it has been assumed that the peak of GW production occurs exactly at the time given by eq. \ref{eq: annihilation condition}, identified with the annihilation time of the network.
As we will see, it is important to determine precisely both the annihilation time of the network and the time of emission of the peak of GWs, 
since the final GW spectrum grows as $\eta^{-4}$, as one can see by extrapolating the scaling properties in eqs.~\ref{eq: rhoA} and~\ref{eq: GW abundance}.

A useful quantity to monitor the degree of annihilation of the network is the fraction of volume occupied by the False Vacuum, $\mathcal{F}_{\text{fv}}(\eta)$.
In a $\mathbb{Z}_2$ model, $\mathcal{F}_{\text{fv}}=1/2$ in scaling. There is some literature on how $\mathcal{F}_{\text{fv}}$  decays during the annihilation phase \cite{Hindmarsh:1996xv,Coulson:1995nv,Larsson:1996sp,Correia:2014kqa, Correia:2018tty} (see also \cite{Pujolas:2022qvs})
with no current consensus.
In the following sections, we will see that the network decays exponentially in the annihilation phase according to
\begin{equation}
\label{eq:Ffv}
\mathcal{F}_{\text{fv}}=0.5 \exp\left[-\left(
\frac{\eta}{\eta_\text{ann}}\right)^p\,\right]
\, ,
\end{equation}
with $\eta_\text{ann}$ possibly differing from $\eta_{\Delta V}$ by $O(1)$ factors, and $p\simeq 3$.

%%%%%%%%%%%%%%%%%%%%%%%%%%%%%%%%%%%%%%%%%%%%%%%%%%%%%%%%%%%%%%%%%%%%%%%

\section{Numerical results}
\label{sec:numerics}

\begin{figure}[t]\centering
	\includegraphics[width=0.5\textwidth]{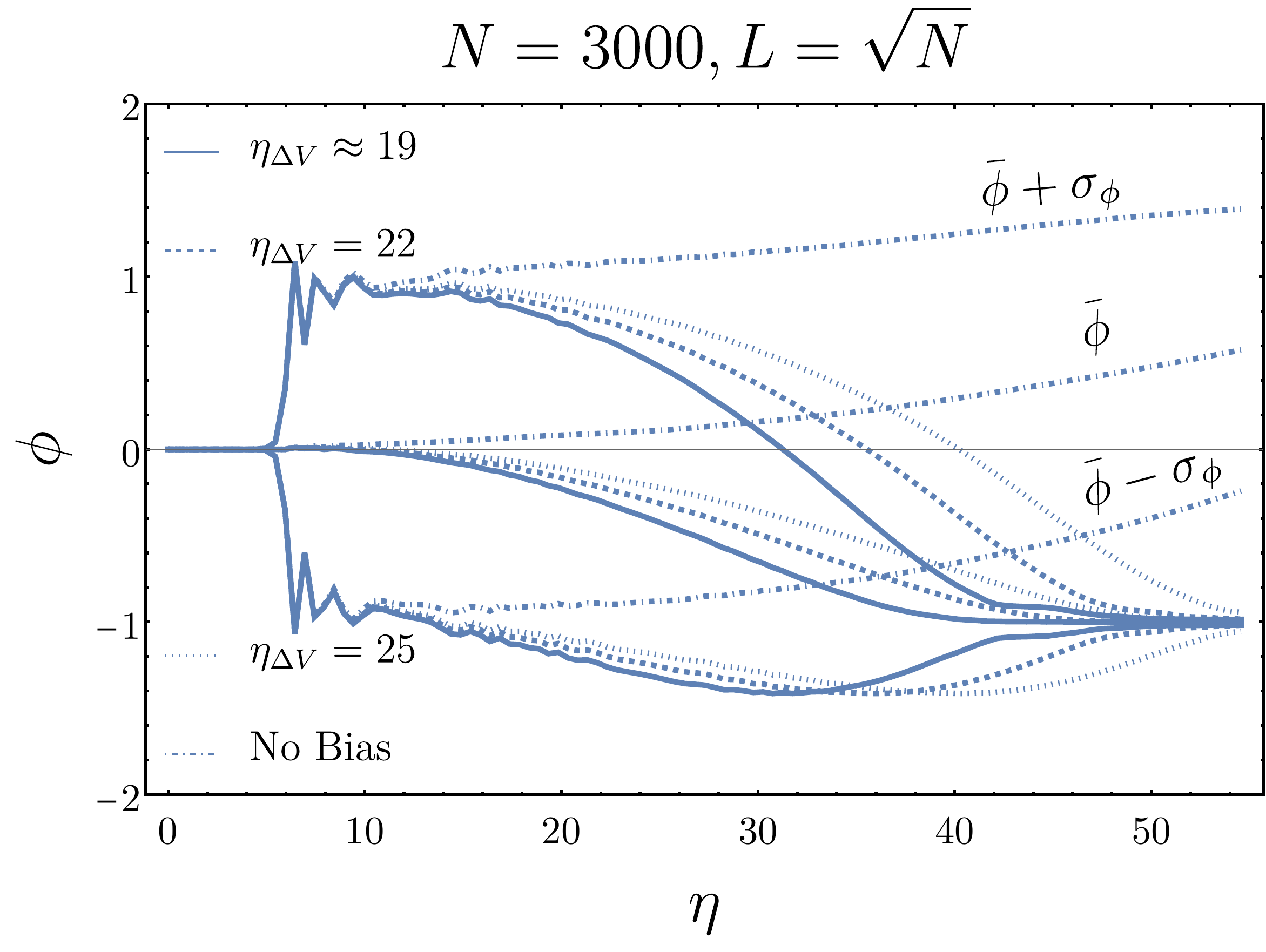}
	\hspace{2em}
	\caption{Evolution of the average of the field in the simulation box, $\bar{\phi}$, plus (minus) its standard deviation $\sigma_\phi$, for several sizes of the bias $\Delta V$,  as a function of conformal time.}
	\label{fig:field}
\end{figure}

We now present the results of our lattice simulations of DW networks, obtained by means of the {\tt Cosmolattice}  code~\cite{Figueroa:2021yhd, Figueroa:2020rrl}. We set initial conditions at the initial conformal time $\eta_i$ in radiation domination such that $m=H(\eta_i=0)$, and $a(\eta_i=0)=1$.   We assign a white noise spectrum of small Gaussian fluctuations in Fourier space to the scalar field, while setting its homogeneous component at $\phi=0$. The number of gridpoints $N^3$ and the comoving box size  $L$ determine the comoving lattice spacing $\Delta x=L/N$. We set $\Delta x$ and the duration $\eta_f$ of our simulations such that the physical lattice spacing $a(\eta_f)\Delta x$ at the end of the simulation is smaller than (or equal to) the domain wall width $\delta_w\sim m^{-1}$, and impose that the simulation box contains at least one Hubble patch at the final time. From here on we set $v=m=1$, which corresponds to the choice $\lambda=1/2$ for the quartic coupling. With these choices, $a(\eta)=1+\eta$, $H(\eta)=(1+\eta)^{-2}$ and the domain wall tension is simply $\sigma =2/3$.

The maximal dynamical range that can be probed with the simulation is then $\eta_{f, \rm max}=\sqrt{N}-1$, obtained by choosing $L=\sqrt{N}$, such that at $\eta_f$ there is only one Hubble patch in the box and $a(\eta_f)\Delta x=\delta_w$. The field evolution is performed using the leapfrog algorithm with a time step $\Delta \eta\lesssim\Delta x/2$, such that the Courant criterion is well satisfied. More details about the numerical setup can be found in Appendix~\ref{app:numerics}. 

Here we shall consider a cubic bias, i.e. $V_{\text{bias}}=q\phi^3$, rather than the linear term. The reason for this choice is only technical: a linear bias shifts the location of the maximum of the potential, thereby introducing a bias in the population of the two minima already at the early times of the simulation. Since the limited dynamical range that can be simulated requires a sizable $\Delta V$, such a population bias would prevent the formation of a network and/or alter its evolution. The cubic bias allows to overcome this problem, as the maximum is not displaced, and for sufficiently small initial field fluctuations  the system does not notice the asymmetry introduced by $V_{\text{bias}}$ in the initial steps.\footnote{An alternative strategy is to use a time-dependent linear bias~\cite{Kitajima:2023cek}, which is initially negligible and becomes important only after a certain activation time. However, we have found that this technique introduces additional uncertainties in the final results, due to different possible choices of the activation time of the bias.} The size of the asymmetry between the two minima is then given by $\Delta V=2q(1+9q^2)^{3/2}$. Therefore, the parameter $q$ determines the time $\eta_{\Delta V}$,  defined by eq.~(\ref{eq: annihilation condition}). 

The scenario relevant to our analysis is that of a domain wall network that achieves the scaling regime sufficiently before it starts collapsing as a consequence of a pressure bias between the two vacua. For the potential adopted in this work, this occurs for $\Delta V/V(0)\lesssim 0.007$, corresponding to $\eta_{\Delta V}\gtrsim 18$ and thus we focus on this range of bias sizes. For larger bias, the network does not fully achieve scaling for a sufficient time before collapse. While this scenario can certainly occur, it is characterized by some residual dependence on initial conditions, which makes it difficult to extract general conclusions.

We show the evolution of the average field value together with its standard deviation as a function of conformal time in Fig.~\ref{fig:field}, as obtained in simulations with a box size $(N=3000)^3$ and maximal time $\eta_f=55$. Here we fixed initial conditions and varied the size of the bias potential, such that $\eta_{\Delta V}=\{19, 22, 25\}$. The following features can be clearly distinguished: first, the field is initially localized very close to the maximum of the potential, until around $\eta\simeq 6$ it relaxes to the two minima. Field oscillations are sizeable until $\eta\simeq 10$, when they are significantly diluted by Hubble friction. The standard deviation then begins to shrink after $\eta\sim 20$ and for $\eta\gtrsim 30$ only the leftmost minimum is populated, signaling that the network is rapidly dissolving under the action of the bias. For comparison, the behavior for vanishing bias is also shown (dot-dashed curve). In this case, a tendency towards the right-most minimum occurs at around $\eta\sim 15$, which is to be attributed to the relatively small (and decreasing) number of Hubble patches at those times, i.e. $\sim (L/\eta)^3\lesssim 50$ at $\eta\gtrsim 15$.

The deviations in the network evolution in the presence of a bias are best understood by focusing on the following quantities. 

\subsection{False Vacuum Fraction}
\label{sec:FVnum}

\begin{figure}[t]\centering
  \includegraphics[width=0.48\textwidth]{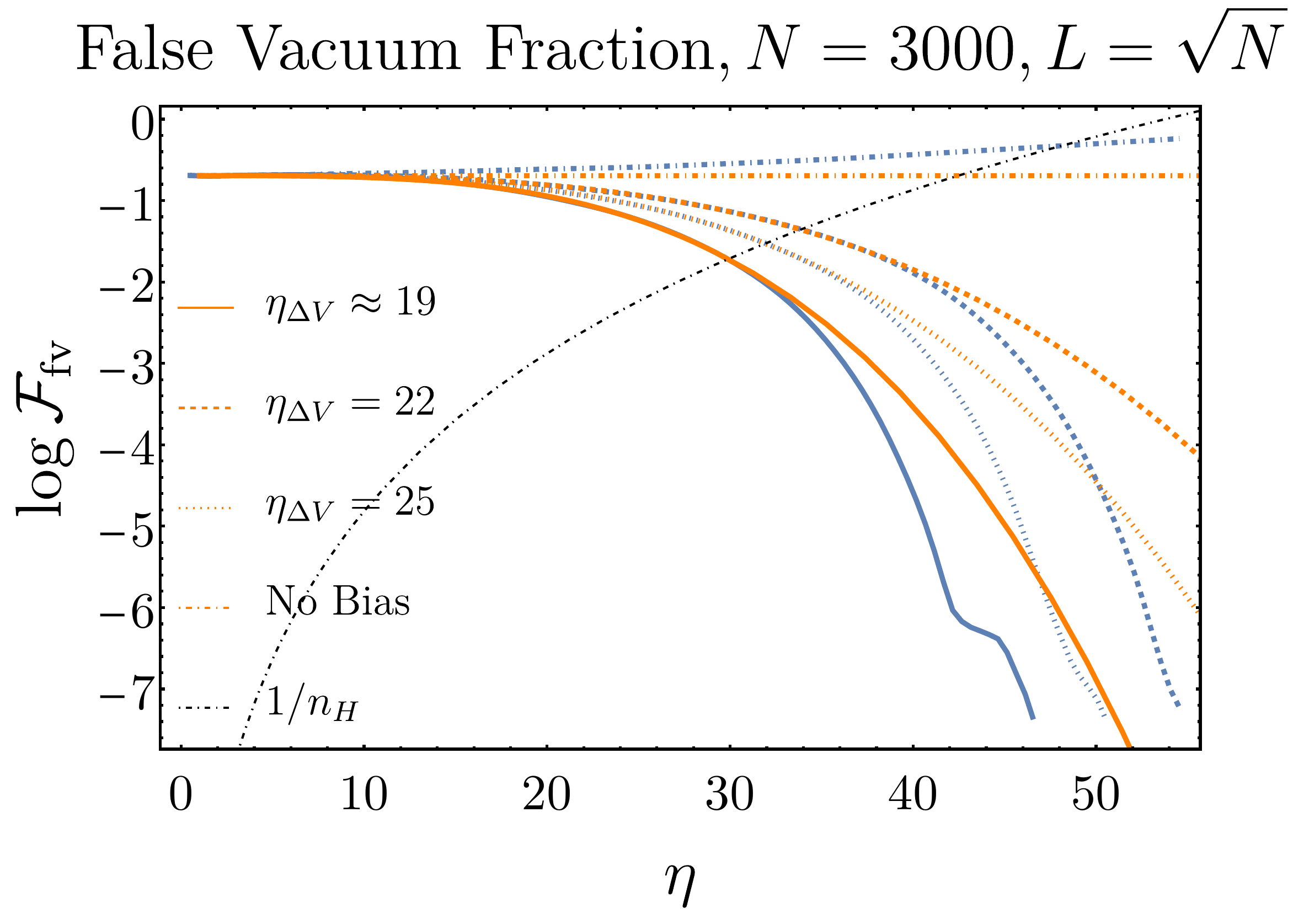}
  \hspace{2em}
  \caption{(Log of) Volume fraction in the simulation box in the false vacuum, as a function of conformal time. The blue curves show the numerical results from our simulations. We fitted such results for small $\eta$, before the curves cross the black dot-dashed curve, that shows the inverse of the number of Hubble volumes in the simulation box.   The orange curves are the resulting fits.}
  \label{fig:fvfraction}
\end{figure}

First, we look at the fraction of volume in the false vacuum $\mathcal{F}_\text{fv}$, which is numerically obtained as the fraction of the simulation grid points where $\phi>0$, shown in Fig.~\ref{fig:fvfraction} (blue curves). As expected, initially $\mathcal{F}_\text{fv}=0.5$ in all simulations. This remains approximately true for the simulation without a bias, although a slight deviation to larger values is observed at late times, corresponding to the aforementioned small number of Hubble patches near the end of the simulation. In simulations where a bias is included, $\mathcal{F}_\text{fv}$ decreases rapidly after a time which depends on the size of the bias, obviously the later the smaller $\Delta V$. In this work, we are mostly interested in false vacuum regions that are at least Hubble-sized, since their radius becomes equal to their Schwarzschild radius if their energy density becomes comparable to the background at Hubble crossing, as we will explain in more detail in Section~\ref{sec:analytics}. 
When the false vacuum fraction drops below the inverse number of Hubble volumes in the box, given by $n_H\simeq (L/(1+\eta))^{3}$ and shown by the black dot-dashed curve in Fig.~\ref{fig:fvfraction}, such regions no longer exist in our simulation. 
For the bias sizes of interest, this occurs around $\eta\simeq 30$. At later times, the remaining false vacuum regions are in sub-Hubble structures. Correspondingly, a much steeper decrease of $\mathcal{F}_\text{fv}$ is observed at late times than at early times, when super-Hubble false vacuum regions can still be present. Thus we attempt fitting the numerical results until the time at which $\mathcal{F}_{\text{fv}}\simeq n^{-1}_H$ with eq.~\eqref{eq:Ffv}
where $\eta_\text{ann}$ and $p$ are fitting parameters. The result of this procedure is shown by the orange curves in Fig.~\ref{fig:fvfraction}. The fitting function above provides an excellent fit to the early time data, and we find $p\simeq 3.3-3.5$. 

\begin{figure}[t]\centering
  \includegraphics[width=0.48\textwidth]{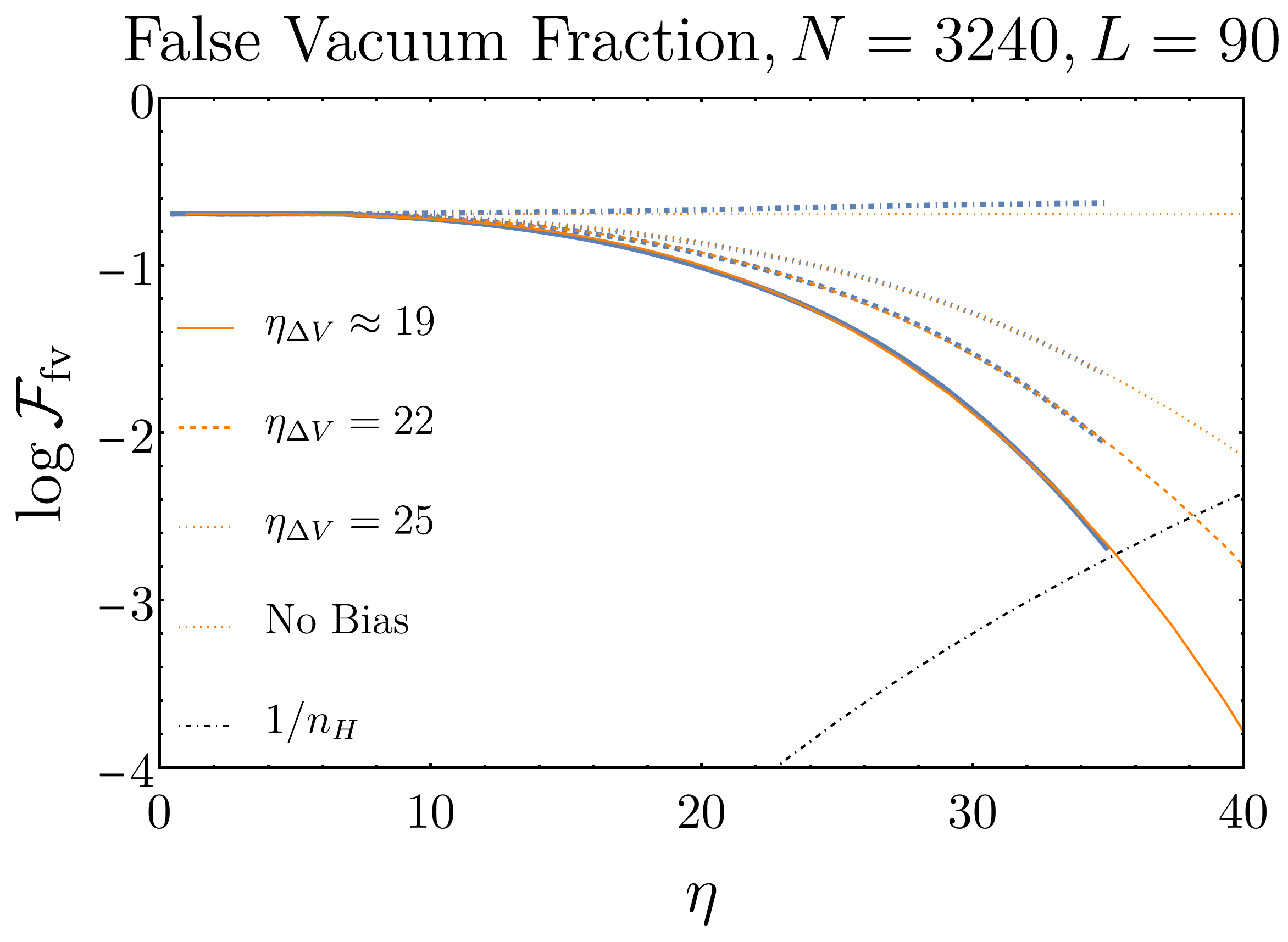}
  \hspace{2em}
  \caption{As in Fig.~\ref{fig:fvfraction}, with $N=3240$ and $L=90$.}
  \label{fig:fvL90}
\end{figure}

To investigate the dependence of these results on the simulation box, we increase the number of Hubble patches in our box by increasing $L$ (and $N$ to a smaller extent) at the price of slightly worsening the spatial resolution, and thus limiting the dynamical range of our simulations. Guided by the previous discussion, we choose $L$ and $\eta_f$ such that $\delta_w/(a\Delta x)\simeq 1$ at time $\eta_f\simeq 35$, corresponding to $L\simeq 90$, with $N=3240$. These choices increase the number of Hubble patches by a factor of $\approx 4.5$ with respect to the results in Fig.~\ref{fig:fvfraction}. The new results are reported in Fig.~\ref{fig:fvL90}, together with the fitting curves. It can be appreciated that $\mathcal{F}_\text{fv}$ now remains almost constant for the entire simulation time in the absence of bias, thereby confirming that the previously observed deviation is due to the limited number of Hubble volumes. The most relevant result of this improved analysis is a decrease of the inferred value of $p$ (which is closer to analytical expectations discussed in Sec.~\ref{sec:analytics}).

We then perform several realizations with increased number of Hubble patches, for several values of bias size and changing the random seed that sets the initial conditions, to estimate numerical uncertainties in our results. We report results in Table~\ref{tab:fitparam}. Averaging over all realizations, we infer
\begin{equation}
p = 3.0\pm 0.3.
\end{equation}
When comparing $\eta_\text{ann}$ to the rough expectation $\eta_{\Delta V}: \sigma H =\Delta V$, we find a slight delay $\eta_\text{ann}\approx 1.5\eta_{\Delta V}$.~\footnote{In this regard, we disagree with \cite{Chang:2023rll} on the dependence of $\eta_\text{ann}$ with $\Delta V$.}

\begin{table}[h]
\centering
\begin{tabular} {| l | c| c|}
\hline\hline
 \multicolumn{1}{|c|}{$\eta_{\Delta V}$} &  \multicolumn{1}{|c|}{$p$} &  \multicolumn{1}{|c|}{$\eta_\text{ann}$} \\
\hline\hline
19.4 & $ 3.0\pm 0.3$ &$ 29.0\pm 0.4$\\
\hline
22 & $3.0\pm 0.1$ & $31.2\pm 0.9$\\
\hline
25 & $3.1\pm 0.3   $ & $33.9\pm 2.5  $\\
\hline
27 & $2.9\pm 0.2 $ & $38.5\pm 1.8$\\
\hline
\end{tabular}
\caption{Mean and standard deviation of the parameters of the fitting function eq.~\eqref{eq:Ffv}, over several realizations of our lattice simulations (about $3-4$ per each row), with different random seeds. }
\label{tab:fitparam}
\end{table}

\begin{figure}[t]\centering
  \includegraphics[width=0.48\textwidth]{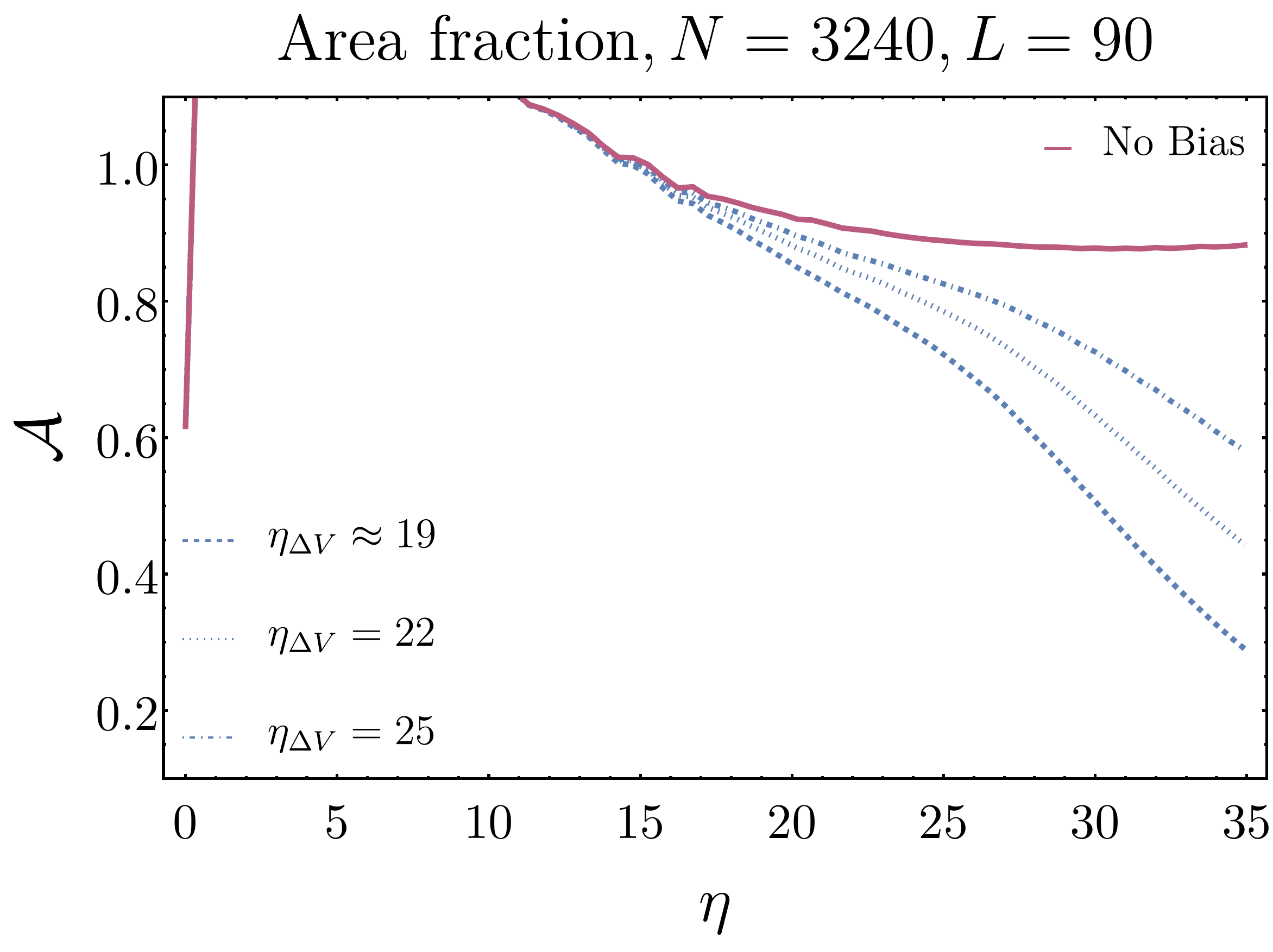}
  \hspace{2em}
  \caption{Area fraction, eq.~(\ref{eq:Area}), as a function of conformal time. Note that at early times, $\eta\lesssim 10$, there is a transient which does not carry physically relevant information.}
  \label{fig:areaparam}
\end{figure}

\subsection{Area parameter}

In the absence of a bias, a domain wall network achieves a scaling regime where the area parameter $\mathcal{A}$ remains approximately constant with time. The energy density in the domain walls during the scaling regime, is determined by eq.~\eqref{eq: rhoA}, which is generally smaller than the total energy density stored by the scalar field. We show in Fig.~\ref{fig:areaparam} the area parameter extracted from our high resolution simulations, in the absence of a bias (solid magenta curve). As expected, it remains approximately constant  for $\eta\gtrsim 25$, saturating for this particular realization to a value $\mathcal{A}\simeq 0.9$. These findings agree roughly with those of~\cite{Hiramatsu:2013qaa}.

The corresponding evolution in the presence of the bias is shown by the blue dashed, dotted and dot-dashed curves, for several bias sizes. One can appreciate that the biased network follows the unbiased one until $\eta\simeq 14-17$, depending on the size of the bias.\footnote{Strictly speaking, the area parameter has not achieved a constant value by that time, therefore simulations with even smaller bias sizes would be desirable to exclude residual effects of initial conditions. With our resources, these can be performed at the cost of lowering the number of Hubble volumes in the simulation box, which alters the determination of the fit parameter $p$. In our high-resolution simulations, we find no significant differences in the inferred value of $p$, nor any significant dependence on initial conditions for $\eta_{\Delta V}\gtrsim 19$, suggesting that the bias sizes explored in this work are sufficient to achieve a scaling behavior at early times, before the bias affects the network evolution.}

\subsection{Energy density}

\begin{figure*}[t]\centering
  \includegraphics[width=0.43\textwidth]{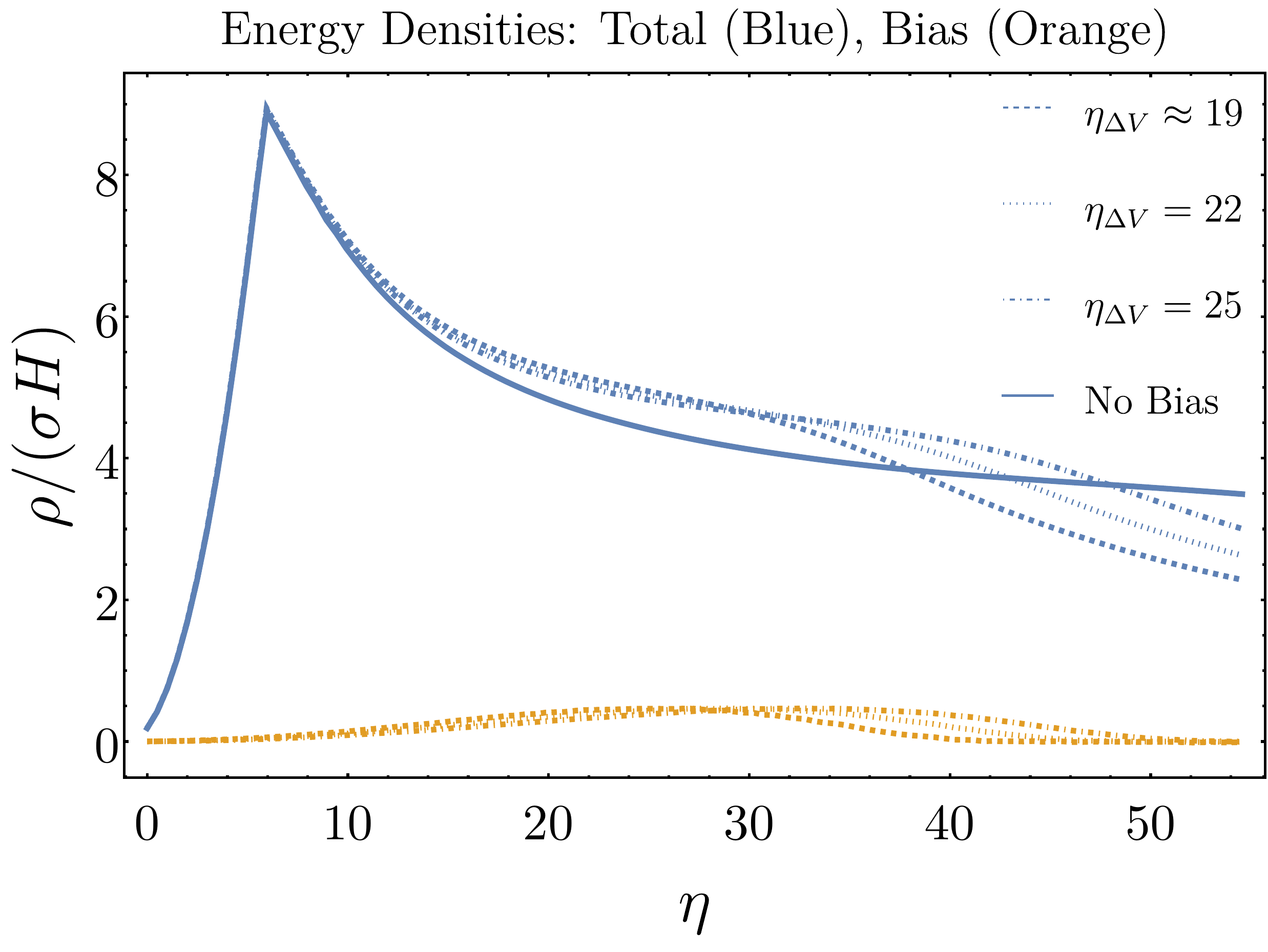}
  \hspace{2em}
\includegraphics[width=0.43\textwidth]{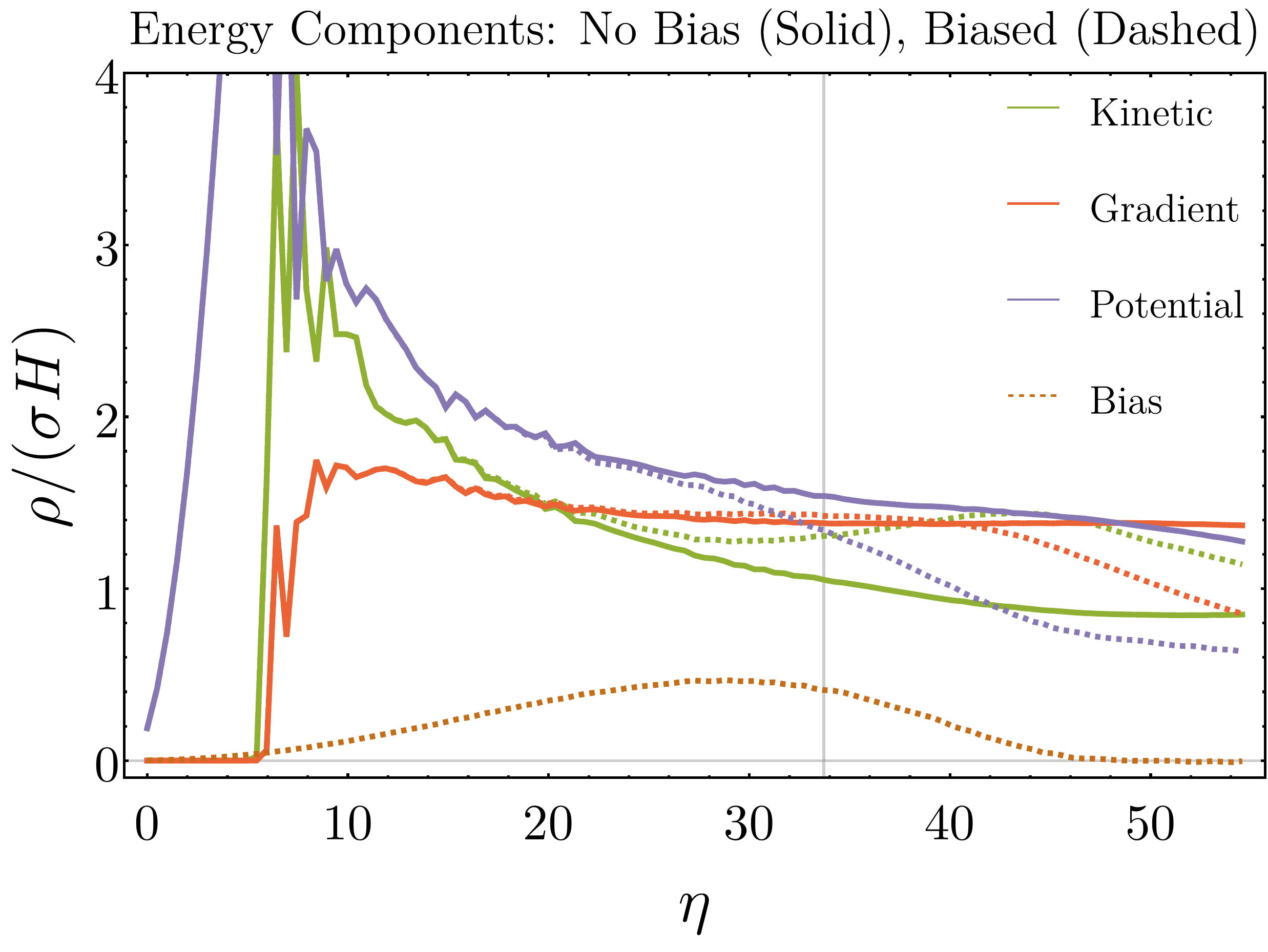}
  \hspace{2em}
  \caption{\emph{Left}: Energy density in the scalar field as a function of conformal time (blue), normalized to the scaling behavior $\sigma H$, without (solid) and with several choices of bias sizes (dashed, dotted, dot-dashed). In orange, the evolution of the average of $V_\text{bias}$ in the simulation box is shown. 
  \emph{Right:}
  Components of the energy densities of the scalar field as a function of conformal time, for unbiased (solid) and biased (dashed) potential, the latter for $\eta_{\Delta V}=22$. The vertical gray line corresponds to the value of $\eta_\text{ann}$ inferred from fitting the false vacuum fraction according to eq.~\eqref{eq:Ffv}. In  both figures, $N=3000$ and $L=\sqrt{N}$.}
  \label{fig:entot}
\end{figure*}

The evolution of the energy density of the scalar field is shown in Fig.~\ref{fig:entot} (left) for unbiased (solid) and biased (dashed, dotted, dot-dashed) potential. These results are obtained in our longest simulations, corresponding to the choice $L=\sqrt{N}$. As expected, in the unbiased case the total energy density in the scalar field remains approximately constant after $\eta\gtrsim 30$, with a final value $\rho_\text{tot}\approx 3.5\, \sigma H$. On the other hand, in the biased case the total energy density decreases sharply, together with the decrease of the vacuum contribution from the bias potential (orange curves), due to the annihilation of the network. The evolution of the three components of the energy density, i.e. kinetic, gradient and potential energy, is shown in Fig.~\ref{fig:entot} (right) for the unbiased (solid) potential as well as for an example case of biased potential (dashed) with the choice $\eta_{\Delta V}=22$. 

The following observations can be made: in the unbiased case, the gradient energy rapidly saturates to a constant value $\rho_\text{grad}\simeq 1.4\, \sigma H$, while the potential energy density $\rho_\text{pot}$ decreases slowly until the end of our simulations (here by potential we denote only the $\mathbb{Z}_2$ symmetric term, whose behavior is shown by the purple curve). The latter decrease may however partially be a numerical effect since it occurs almost at the end of the simulation where the domain wall width becomes comparable to the lattice spacing of the simulation. Overall, these two components make most of the energy density of the scalar field, in agreement with the intuition that most of the energy density is in domain walls, and in particular $\rho_\text{grad}+\rho_\text{pot}\simeq 2.6 \, \sigma H$ at the end of the simulation. On the other hand, the kinetic energy decreases rapidly until it saturates to an approximately constant value $\rho_\text{kin}\simeq 0.9\, \sigma H$, thereby making a subleading contribution to the energy density. 

The situation in the biased case is strikingly different, where deviations from the unbiased scenario occur around $\eta\gtrsim 20$: most importantly, the kinetic energy stops decreasing and quickly begins to increase, while the potential energy decreases sharply. The former overcomes the latter around $\eta\gtrsim 34$. This is very close to the value of the annihilation time $\eta_\text{ann}$ inferred from our fit of the false vacuum fraction, nicely confirming that this time scale indeed characterizes the annihilation of the network when the vacuum pressure from the bias term (shown by the dashed brown curve) accelerates the walls, thereby increasing the kinetic energy. The gradient energy initially remains constant, before sharply decreasing at $\eta\gtrsim 40$. This is easy to interpret, as the existence of the network is the source of gradient energy, which is thus quickly dissipated away once the walls annihilate. Eventually, at $\eta\gtrsim 45$ also the kinetic energy starts decreasing. We have checked that for $\eta\gtrsim 45$ both the potential and kinetic components decrease approximately like non-relativistic matter, as expected since we are working with a massive scalar field with $m\gg H$ at the end of the simulation. Overall, the kinetic energy  dominates the energy density of the biased domain wall network at the end of our simulations. Notice also that the bias term very rapidly vanishes after $\eta\gtrsim 40$, corresponding to the exponential decay of false vacuum regions.

\begin{figure}[t]\centering
  \includegraphics[width=0.48\textwidth]{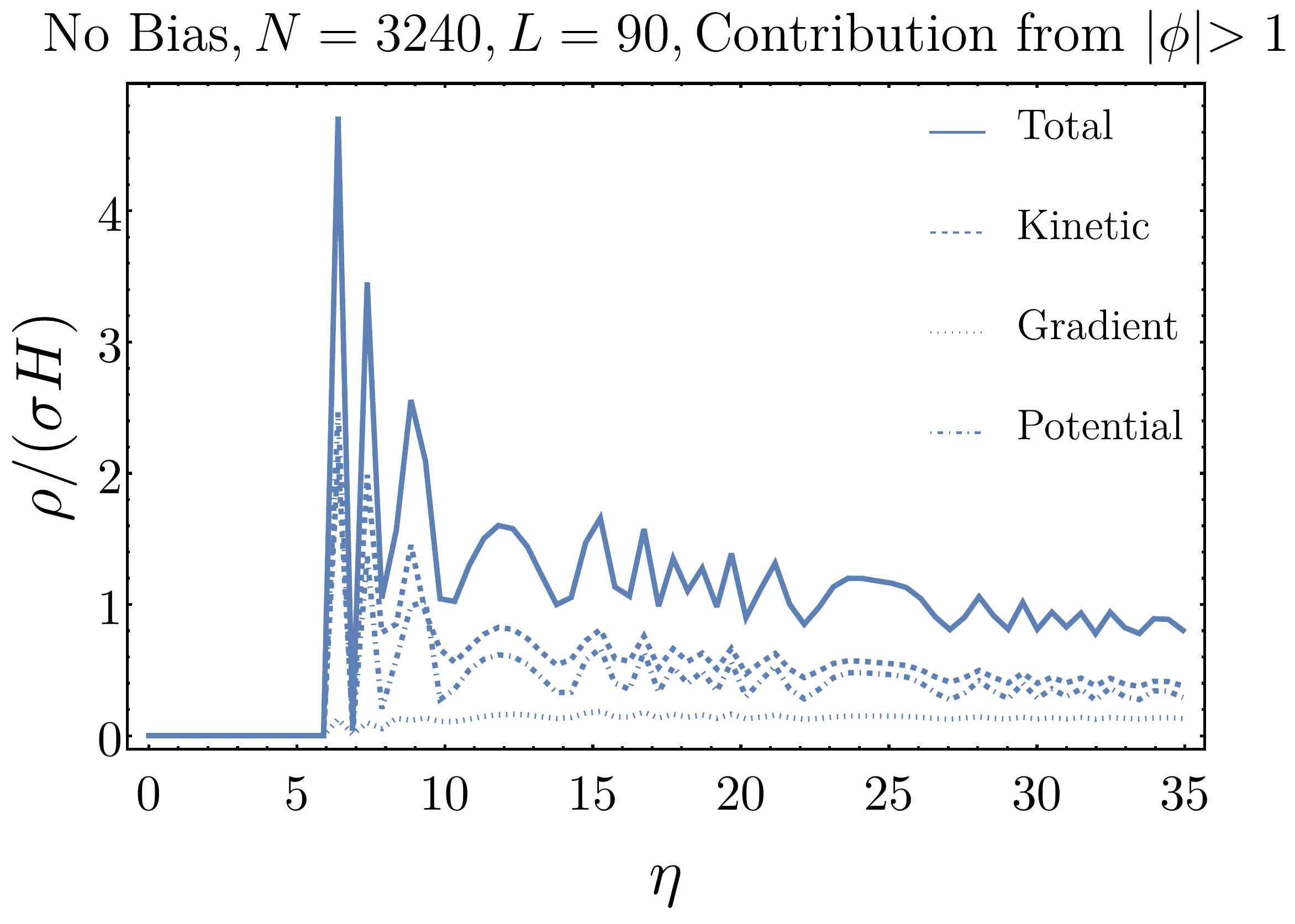}
  \hspace{2em}
  \caption{Energy density in the simulation box corresponding to $\lvert \phi\lvert>1$. Here $N=3240$ and $L=90$.}
  \label{fig:enwaves}
\end{figure}

Our findings on the behavior of the energy density in the unbiased case may seem different from the common lore in the literature, which mostly adopts $\rho_{\text{dw}}\simeq 2\mathcal{A}\sigma H$. In our simulations, we find that the total energy density in the scalar field, including all the simulation box, is roughly twice as large as the estimate above, $\rho_\text{tot}\approx 3.5\sigma H$. One should nonetheless notice that: 1) the commonly adopted estimate is expected to apply only to static domain walls, i.e. it neglects the contribution of the kinetic energy, which in our simulations accounts for $\rho_\text{kin}\approx 0.9\sigma H$; 2) the simulation box includes regions where $\lvert\phi\lvert>1$, which cannot be attributed to domain walls, but rather to scalar waves. The energy density in this region of field space is reported in Fig~\ref{fig:enwaves}, for a simulation with a large number of Hubble patches. Its size at the end of the simulation is $\rho_\text{scal}\approx 0.8\sigma H$, mostly coming from kinetic and potential energy. Ignoring the kinetic part, it contributes $\approx 0.5\sigma H$. Therefore our larger total energy density in the scalar field is easily explained in terms of the two contributions above (plus a small unavoidable contribution from scalar waves in the region $\lvert\phi\lvert\leq 1$).\footnote{Despite the commonly reported estimate above, we notice that our findings agree with those presented in~\cite{Hiramatsu:2013qaa}.} We conclude that, according to our simulations, domain walls carry an energy density $\rho_{\text{dw}}\simeq 2.4\sigma H$. If interpreted in terms of a relativistic correction to the standard formula, i.e. $\rho_{\text{dw}}\simeq 2\mathcal{A}\gamma^2\sigma H$, it implies $\gamma\approx 1.2$.

\subsection{Gravitational Waves}

It has long been appreciated that a domain wall network acts as an efficient source of GWs during its evolution. Previous numerical calculations have mostly focused on the contribution from the scaling regime. In our simulations, we are able to extract the energy density radiated in GWs throughout the annihilating phase as well. Due to higher memory consumption, our simulations including GWs are limited to $N=2040$, with a maximal simulation time $\eta_f\lesssim 45$. Additionally, to speed up the calculation, we only start the numerical computation of the GWs at the latest times ($\eta>35$). This is justified since we find that, unlike previous estimates, most of the GWs are emitted during the annihilation epoch rather than in the scaling regime (see also~\cite{Kitajima:2023cek}).

\begin{figure}[t]\centering
  \includegraphics[width=0.48\textwidth]{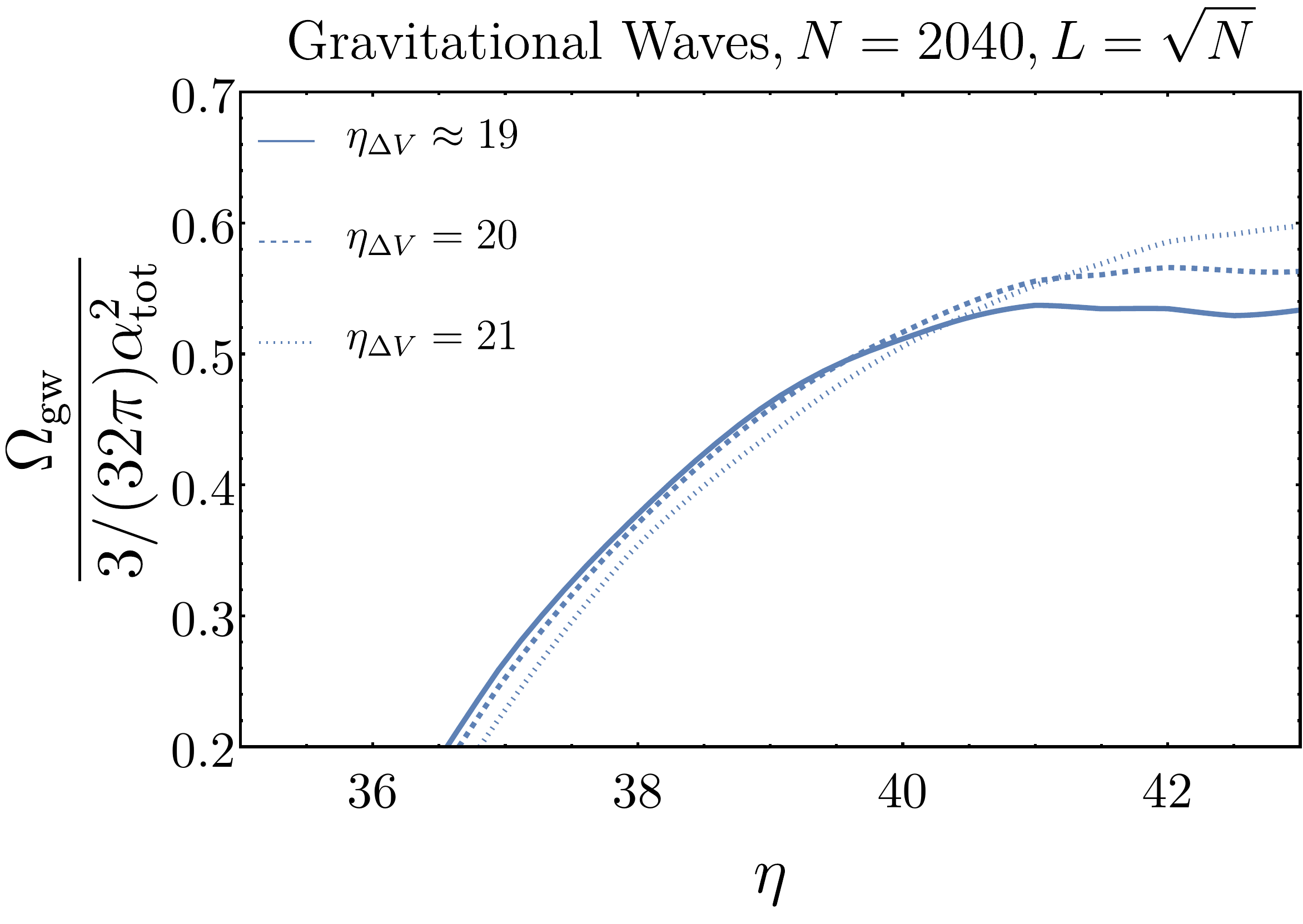}
  \hspace{2em}
  \caption{Energy density fraction in GWs compared to quadrupole formula estimate from the total energy density available in the scalar field. 
 Note that the initial $\Omega_{gw}$ is very small because we start the computation of GWs only at $\eta=35$ for computational reasons.  }
  \label{fig:gw1}
\end{figure}

\begin{figure}[t]\centering
  \includegraphics[width=0.48\textwidth]{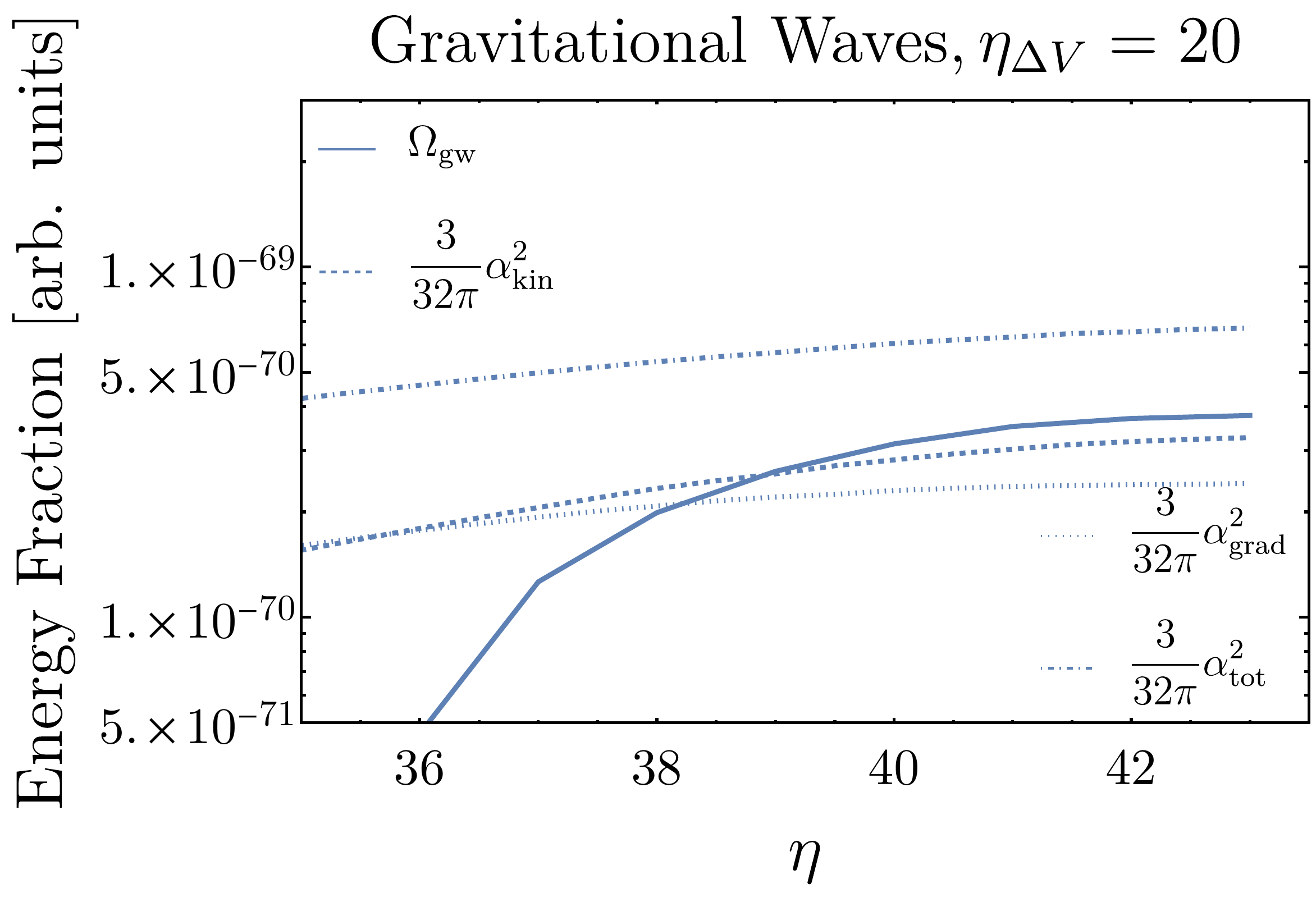}
  \hspace{2em}
  \caption{Energy density fraction in GWs compared to quadrupole formula estimate from the total energy density available in the scalar field.}
  \label{fig:gw2}
\end{figure}

A simple estimate of the maximal energy density fraction (i.e.~the energy density in GWs compared to that of the radiation background) is provided by the quadrupole formula, which gives $\Omega_\text{gw}(\eta)\sim 3/(32\pi)\alpha^2_\text{tot}(\eta)$, where $\alpha_\text{tot}\equiv \rho_\phi/(3 H^2 M_p^2)$. This would correspond to the case in which all the energy density in the scalar field sources GWs. We compare our results with this simple estimate in Fig.~\ref{fig:gw1}, for three different choices of bias size for which we are able to capture the full GW production (with same initial conditions as in all previous figures). For all our choices, we find that the GW energy density fraction reaches a peak at $\eta\gtrsim 40$. The efficiency factor with respect to the simple quadrupole estimate at peak production is $\epsilon\equiv \Omega_\text{gw}/(3/(32\pi)\alpha^2_\text{tot})\simeq 0.5-0.6$. Fig.~\ref{fig:gw2} shows that GWs are compatible with being mostly sourced by the kinetic component of the energy density in the scalar field, with a subdominant contribution from the gradient component, for one example value of bias size. The GW energy density fraction is computed by integrating the numerically obtained GW spectrum, although the final result is dominated by the region around the peak of the spectrum.

Our results importantly point to a mild delay between the characteristic time of the annihilation epoch $\eta_\text{ann}$ and the time at which most GW production occurs, $\eta_\text{gw}$. In our simulations, this is estimated to be 
$$\eta_\text{gw}/\eta_\text{ann}\simeq 1.3-1.4 \, .$$ 
Physically, our findings suggest that efficient GW production continues throughout the annihilation epoch, and so beyond the production during scaling studied in previous works, 
and that most of the relic abundance of GWs is determined by the late stages of the DW network collapse.
Compared to the previous literature, the delay amounts to a factor of $~1.3\times 1.4 \sim 1.8$ in the time of GW emission, which was previously estimated to be $\eta_{\Delta V}$, and so, following eq.~(\ref{eq: GW abundance}), an enhancement of $1.8^4 \sim 10$ in the total GW abundance.

\section{Semi-Analytical approach} 
\label{sec:analytics}

Let us now turn to a different way to analyze the problem. The network annihilation process can be viewed as a transition from DWs in the scaling regime to, eventually, a collection of rare FV pockets. During scaling, in each Hubble patch, there is typically one Hubble-sized DW, some scalar radiation and, to a good approximation, no closed DWs. Most of the energy is stored in the form of large DWs and any DW is typically separated from neighbour DWs by the correlation length, which in scaling is set by the Hubble length. 

\begin{figure}[t]\centering
  \includegraphics[width=0.48\textwidth]{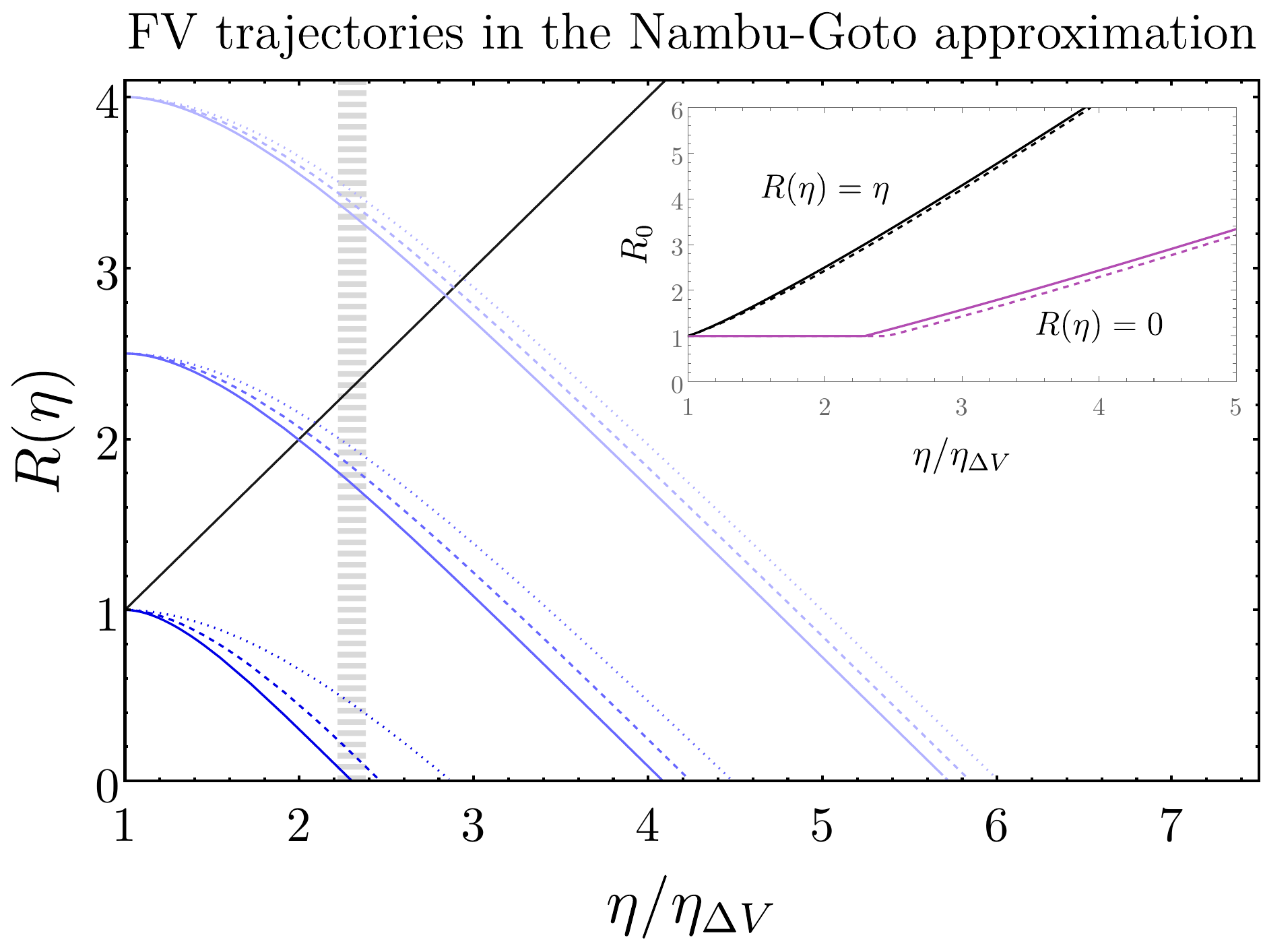}
  \hspace{2em}
  \caption{Evolution of FV pockets with spherical (solid), cylindrical (dashed) and planar (dotted) shapes as extracted from the Nambu Goto approximation (see Appendix~\ref{section:NG}). 
  Blue lines show the comoving radius $R(\eta)\equiv R(\eta;R_0)$ as a function of conformal time for various initial radii $R_0$, starting at $\eta_{\Delta V}$ from rest. 
  Lower $R_0$ pockets are more numerous.
  The dashed vertical gray line denotes when most of the pockets shrink to small sizes. This is expected to coincide with the peak of GW production $\eta_\text{gw}$. 
  The black solid line is the comoving Hubble length at each time. 
The inset shows the initial values of $R_0$ of (spherical or cylindrical) pockets that reach zero (purple) or the Hubble length  (black) at a given time. }
  \label{fig:traj}
\end{figure}

Annihilation starts when the force per unit area from $\Delta V$ is larger than $\sigma H$, pushing the DWs to reduce the volume in the FV. Let us assume that this effect turns on instantaneously at $\eta_{\Delta V}$. Since the typical separation between walls is also of order $\eta_{\Delta V}$, a Hubble-sized FV pocket takes a time of about $\eta_{\Delta V}$ to shrink to zero, and so one expects that 
the fraction of the volume in FV becomes tiny after a delay of order $\eta_{\Delta V}$, that is, at $\eta\simeq 2\eta_{\Delta V}$, since after that time only very rare structures survive, i.e, the ones that started with super-Hubble size at $\eta_{\Delta V}$. As an additional confirmation beyond the numerical results shown in section~\ref{sec:FVnum}, this delay can be qualitatively reproduced by solving the equations of motion for a DW enclosing the FV pocket, in the so-called Nambu-Goto approximation, for some simple DW shapes (see Appendix~\ref{section:NG} for details), as shown in  Fig.~\ref{fig:traj}.

This simple observation has two interesting consequences. First, we identify that  $\eta\sim 2\eta_{\Delta V}$ is a `maximal shrinking time', when most of the FV has shrunk to zero, and so it is natural to expect an additional contribution to GW production on top of the GW that the DWs have sourced during scaling. Qualitatively, this confirms the results presented in the previous section of $\eta_\text{gw}\sim 2\eta_\DV$.

Second, this also sets the time when we can start to picture the `remainders' of the network in a simple and useful way: as an ensemble of FV pockets of different sizes, which are placed far apart so that we can treat them independently. One can call this a dilute gas of FV pockets.

\subsection{False Vacuum fraction}
\label{sec:FVan}

This approximation allows to compute the FV volume fraction ${\cal F}_\text{fv}$ and extrapolate it to times that are inaccessible with numerical methods. The basic idea is simple: FV pockets shrink in time (see Fig.~\ref{fig:traj}), so that the lifespan of each pocket is determined by its initial size $R_0$ at $\eta=\eta_\DV$. Once the network is sufficiently fragmented 
we can approximate ${\cal F}_\text{fv}(\eta)$ as a sum over an ensemble of pockets of different initial radii $R_0$.  Moreover, the relative weights in the sum  $R_0$ are known because they are inherited from the scaling regime.

Indeed, since there are only 2 vacua (for a $\mathbb{Z}_2$ model) that are essentially distributed randomly during scaling, the probability of finding a super-Hubble region of radius $R_0$, where the field is in one vacuum, is expected to be
\begin{equation}\label{P0}
P_0(R_0) =  2^{-(R_0/L_\ann)^3}  ~,    
\end{equation}
with $L_\ann$ the correlation length at the onset of annihilation, which is identified as
$$
L_\ann\simeq \eta_\DV\,.
$$ 
The distribution eq.~\eqref{P0} satisfies the wanted normalization: initially, a Hubble-sized region with $R_0=L_\ann$, has $50\%$ chance to be in either vacuum.

The FV volume fraction then can be obtained by adding up the (shrinking) volumes of all pockets  with weights given by eq.~\eqref{P0} over $R_0>L_\ann$,
\begin{equation}\label{Sint}
{\cal F}_\text{fv}(\eta) = \frac{{\cal N}}{L_\ann^3} \, \int_{L_\ann}^\infty P_0(R_0) \; R^3(\eta;R_0) \; \frac{d R_0}{R_0} \, .
\end{equation}
 We assume here a flat integration measure in $\log R_0$ for simplicity, but the results do not depend very dramatically on the measure choice. Even if this formula holds only for $\eta>2\eta_{\Delta V}$, the overall normalization constant ${\cal N}$ can be fixed to have ${\cal F}_\text{fv}=1/2$ when extrapolated to initial time $\eta_{\Delta V}$.

\begin{figure}[t]\centering
  \includegraphics[width=0.48\textwidth]{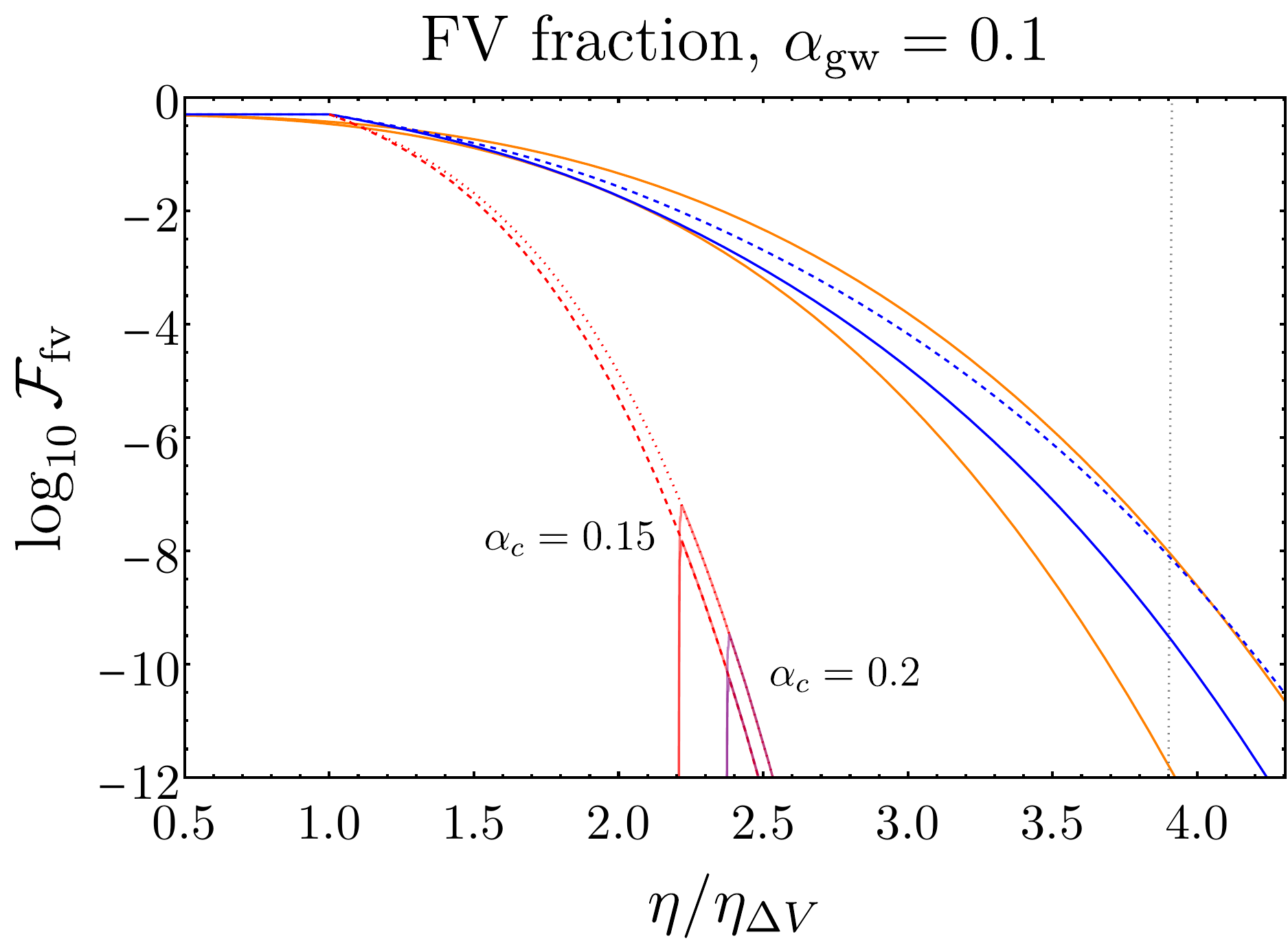}
  \hspace{2em}
  \caption{ FV fractions from numerical simulations and from analytic estimates as a function of conformal time. Orange lines correspond to eq. \eqref{eq:Ffv} with $\eta_\ann$ and $p$ extracted from the fits to the simulations shown in Table~\ref{tab:fitparam} (only the first and third row, giving the lowest and highest fractions respectively). The remaining curves arise from the analytic FV pocket `gas' approximation. Blue lines are the estimate of the full FV fraction from eq. \eqref{Sint} with $w=0.9$ (solid) and $0.8$ (dashed), which represents a justified margin of uncertainty (see Fig.~\ref{fig:collTime}).  The red curves are two estimates of the FV fraction contained in Hubble-sized (or larger) pockets, for a spherical (dashed) and a cylindrical (dotted) pocket. For $\alpha_\text{c}=1$ these correspond to the intersection of the red curves with the dotted vertical line - a tiny number well outside the plot range.  The FV fractions for other choices of the  effective collapse criterion $\alpha_c$ are given by the intersections of the vertical lines and the red curves. 
   }       
  \label{fig:FVfracs}
\end{figure}

Generically, pockets shrink and vanish after some time. The `trajectories' $R(\eta;R_0)$, that we obtain by solving the dynamics in the Nambu-Goto approximation (see Appendix \ref{section:NG}) are basically triangular, they decrease to zero and remain zero. As shown in  Fig.~\ref{fig:traj}, the shrinking time is quite shape independent too. At any time $\eta$, one can track back the initial radius of the pocket that reaches $R=0$ at that moment. We call this 
\begin{equation}\label{R0min}
R_0^\text{min}(\eta)  \,: \quad  R(\eta;R_0^\text{min})=0 \, ,
\end{equation}
and show it in the inset of Fig.~\ref{fig:traj} (purple curves)\footnote{Since there is a slight shape dependence, we take two values of $R_0^\text{min}(\eta)$, from the spherical and the cylindrical pockets, to give a sense of `theoretical' errors.}. By construction, then, the lower integration limit in eq.~\eqref{Sint} can be replaced by $R_0^\text{min}(\eta)$. Since the size distribution in eq.~\eqref{P0} is exponentially biased towards small $R_0$, it is clear that  ${\cal F}_\text{fv}$ must be suppressed by $2^{-(R_0^\text{min}/\eta_\DV)^3}$. 
The Nambu-Goto approximation also provides some useful information to narrow down the asymptotic behaviour at large $\eta$. Figures \ref{fig:traj} and \ref{fig:collTime} suggest that the mock trajectory $R(\eta;R_0) \to R_0 - w (\eta-\eta_\DV) $ with $w$ an $O(1)$ constant gives a reasonable approximation. With this, the asymptotic form
\begin{equation}\label{Sasy}
{\cal F}_\text{fv}\sim 
\left(\frac{\eta_\DV}{\eta}\right)^9\exp\left[- \log2\,\left(\frac{w\,\eta}{\eta_{\DV}}\right)^3 \right] \, ,
\end{equation}
follows. Ignoring the power law term, this allows to recognize the FV decay time introduced in eq.~\eqref{eq:Ffv} as
\begin{equation}
    \eta_\ann = \frac{\eta_\DV}{(\log 2)^{1/3}\, w}\,.
\end{equation}
The value of $1/w$ can be read off from Fig.~\ref{fig:collTime}, which in the end results in $\eta_\ann / \eta_\DV$ being around $1.3$. This is in quite remarkable agreement with the numerical simulations, see Table~\ref{tab:fitparam}, representing an important nontrivial validation of the FV pocket picture. This is also manifest in Fig.~\ref{fig:FVfracs}, where we compare the fits from the numerical simulations (orange curves) to the analytic expression eq.~\eqref{Sint} (blue curves).

In this picture, it is also possible to isolate the part of this FV fraction ${\cal F}_\text{fv}^\text{hor}$ given by Hubble sized pockets (or larger) at any time. It  reduces to integrate from a higher value of $R_0$, the one corresponding to pockets that enter the Hubble radius (rather than vanishing) at that time. We then introduce
\begin{equation}\label{R0hor}
R_0^\text{hor}(\eta)  \,: \quad  R(\eta;R_0^\text{hor})=\eta \, ,
\end{equation}
which is also shown in the inset of Fig.~\ref{fig:traj} (black line). Notice how this radius grows significantly faster than $R_0^\text{min}(\eta)$, hence the part of the FV fraction in Hubble sized (or bigger) pockets is much smaller than the total ${\cal F}_\text{fv}$.

In this case we obtain
\begin{equation}\label{Shor}
{\cal F}_\text{fv}^\text{hor}
   \sim 
   2^{- \left(R_0^\text{hor}(\eta)/ \eta_{\DV}\right)^3} \, .
\end{equation}
Since  $R_0^\text{hor}$ grows linearly in $\eta$ (see Fig.~\ref{fig:traj}), we conclude that the FV fraction contained in super-Hubble pockets behaves asymptotically like eq.~\eqref{eq:Ffv} with $p=3$.

Note also that a very simple relation holds if we use the approximate mock trajectory linear in $\eta$ given above, i.e. $R_0^\text{hor}(\eta)=R_0^\text{min}(\eta)+\eta$, which immediately implies that the fraction in Hubble-sized patches  is:
${\cal F}_\text{fv}^\text{hor} \sim P_0(R_0^\text{hor}) \sim 2^{-((1+w) \eta/\eta_{\Delta V})^3}$ with $w\approx 0.8$, much more suppressed than the total fraction in FV pockets as can be appreciated in Fig. \ref{fig:FVfracs}.
while the total fraction in FV pockets is roughly  ${\cal F}_\text{fv} \sim P_0(R_0^\text{min})\sim 2^{-(w \,\eta/\eta_{\Delta V})^3}$.

These results are reminiscent of the scaling $\exp (-\eta^d)$, in $d+1$  dimensions, suggested by~\cite{Hindmarsh:1996xv}. However, let us emphasize that the similarity is accidental. The analysis of~\cite{Hindmarsh:1996xv} refers actually to an annihilation mechanism based on population bias, with $\Delta V=0$. In $2+1$ dimensions, it was found in \cite{Correia:2018tty} that the analog of \cite{Hindmarsh:1996xv} actually did not apply for the population bias case, whereas it did work for pressure bias, with $\Delta V\neq0$. Our work extends the analysis of~\cite{Correia:2018tty} to $3+1$ dimensions (see also~\cite{Chang:2023rll}) and clarifies the physical reasons behind this decay law for the pressure bias ($\Delta V \neq 0$) mechanism.

\subsection{PBH formation}

We are now ready to give some rough estimates of the abundance of PBHs produced during the network decay. As in~\cite{Ferrer:2018uiu}, the natural strategy is to follow the various FV pockets as they shrink. Part of this evolution takes place while the pocket is super-Hubble sized. 
It is useful then to look at the `figure of merit' defined by the ratio of the Schwarzschild radius of the pocket to the Hubble radius when the pocket size crosses the Hubble radius. This quantity actually coincides with the local overdensity of a FV  with energy density $\rho_\text{pocket}$, 
\begin{equation}
    \alpha_\text{loc} = \frac{\rho_\text{pocket} } {3 M_P^2 H^2}\,.
\end{equation} 
 For $\alpha_\text{loc}\ll 1$ the pocket needs to contract significantly after entering the Hubble radius, in order to form a PBH. This is less likely to happen if it is non-spherical and, since FV pockets descend from a DW network,  asphericities can actually be large. 

For larger $\alpha_\text{loc}$, this is not so challenging and so one can expect PBHs to form  in this way. Moreover,  $\alpha_\text{loc} $ grows in time, so some PBHs are certainly produced.  Notice that the limit $\alpha_\text{loc} \to 1$ is special: the pocket collapses to a BH as soon as it enters the cosmological Hubble radius. These BHs are actually expected to carry a baby-universe. For spherical symmetry, they have been considered in \cite{Garriga:2015fdk,Deng:2016vzb,Blau:1986cw,Berezin:1982ur}. We will consider both types of BHs, but we anticipate that  baby-universe BHs should be much rarer than the ordinary ones. 

The rest of the argument to estimate the PBH abundances is as follows. First, the overdensity produced by  FV pockets that enter the Hubble radius  at $\eta$ scales like 
\begin{equation}\label{alphaloc}
\alpha_\text{loc} \approx 1.5\, \alpha_\text{gw} \left( \eta/\eta_\text{gw}\right)^4   \, , 
\end{equation}
with $\alpha_\text{gw}$ the average fraction of energy density in DWs at $\eta_\text{gw}$, and the prefactor fixed by numerical simulations, see  Appendix~\ref{section:NG}.

Second, as argued before, the collapsing network can only be acceptably approximated as an ensemble of pockets after $\eta\gtrsim 2 \eta_\DV$.

Third, in principle one should do an analysis of how many of the different pockets actually manage to shrink enough to form BHs. This will depend on their degree of asphericity and angular momentum and it can be model dependent. Its estimate deserves a dedicated analysis of data from numerical simulations that is outside the scope of this work. The result of such an analysis should effectively result in a threshold of collapse, $\alpha_c$, such that (on average) pockets which reach $\alpha_\text{loc} \geq \alpha_c$ collapse into a BH.

At present we are unable to obtain a reliable estimate of $\alpha_c$. We thus consider a range of benchmark values that could be reasonable for this quantity. Since $\alpha_c=1$ is the threshold for baby-universe BHs, $\alpha_c<1$ corresponds to sub-Hubble BHs.

Fourth, we will identify the PBH abundance (of either type) with the  FV fraction ${\cal F}_\text{fv}^\text{hor}$ evaluated at the time $\eta_\text{PBH}$ when $\alpha_\text{loc}  = \alpha_c$ is met.

We show in Fig.~\ref{fig:FVfracs} the FV fraction expected to collapse into BHs according to this simplified criterion, with $\alpha_c = 0.15$ and $0.2$. 
Clearly, since the collapse criterion is satisfied first with smaller $\alpha_c$, the sub-Hubble PBHs are much more abundant than the ones carrying baby universes.  

As expected, the abundance is exponentially sensitive to $\alpha_c$. 
The abundance of  baby-universe BHs ($\alpha_c\simeq1$) is extremely suppressed (and given by the extrapolation of the red curves to the vertical dotted line) even for quite large $\alpha_\text{gw}$. On the other hand, the sub-Hubble PBHs that can be formed `soon' could be  much more abundant. The basic reason why sub-Hubble PBHs are more abundant is simply that the FV pockets they descend from are (extremely) more common.
Unfortunately, with the current analysis, it is difficult to make any further quantitative statements due to the large uncertainty in $\alpha_c$. A more detailed study is left for the future.

\section{Phenomenological Implications}
\label{sec:pheno}

\begin{figure}
    \centering    \includegraphics[width=0.47\textwidth]{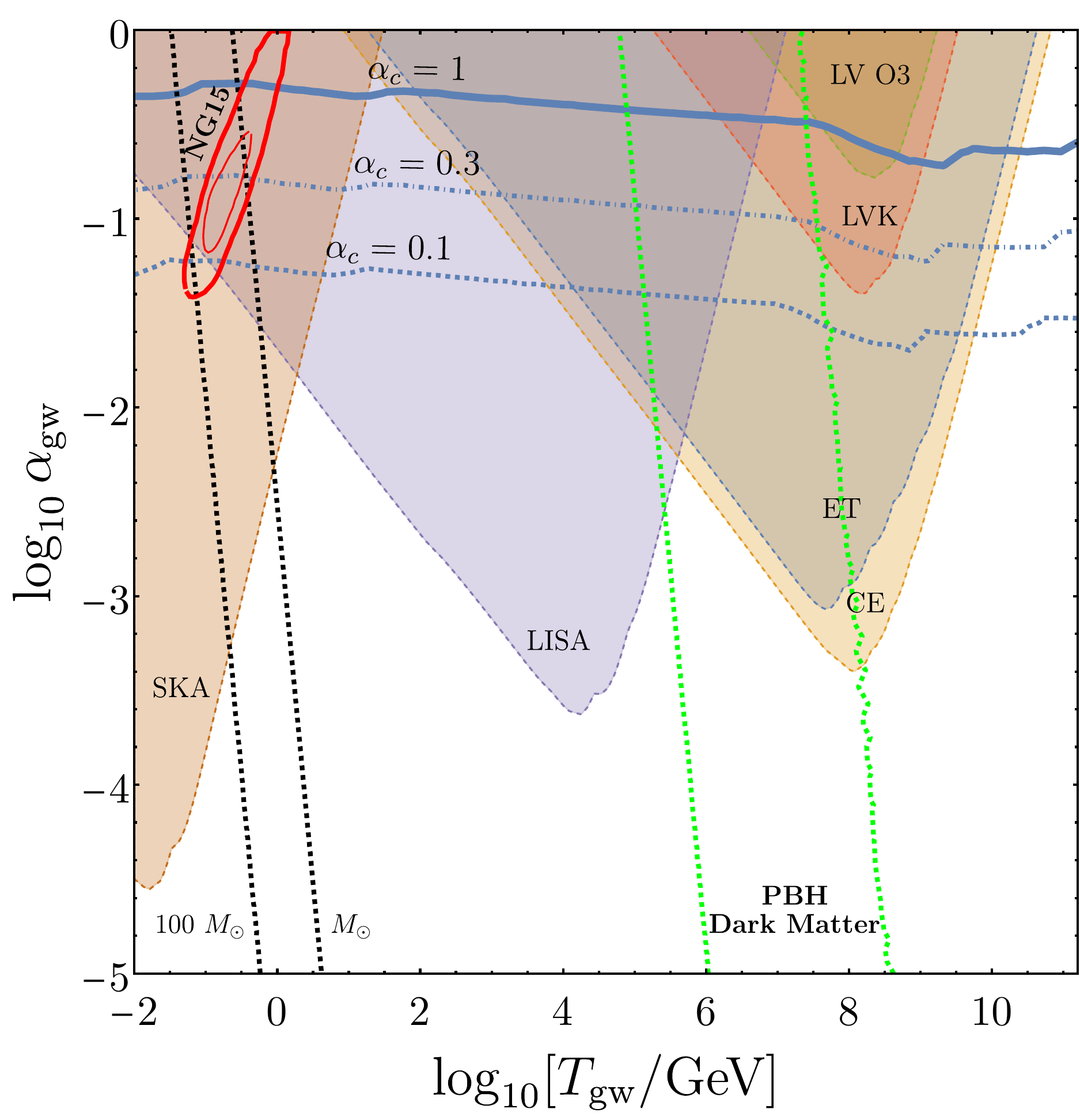}
    \caption{Constraints on the PBH abundance from the collapse of the DW network for different values of the collapse threshold $\alpha_c$ (blue lines) in terms of the fraction of the energy density in the DW network at the time of GW emission ($\alpha_\text{gw}$) and the background temperature at that point ($T_\text{gw}$). The region above the solid blue curve is conservatively excluded. The red ellipsis shows the region of parameters where the DW network interpretation explains the recent GW signals detected at PTA \cite{NANOGrav:2023hvm} at one and two standard deviations. The shaded regions are the constraints on the GW spectrum from LIGO-VIRGO O3 data (LV) and the prospects for detection with LIGO-Virgo-KAGRA design sensitivity (LVK), Einstein Telescope, Cosmic Explorer, LISA and SKA using the power-law integrated sensitivity curves from \cite{Schmitz:2020syl}. Finally, the green band corresponds the asteroid mass range ($10^{-16}-10^{-11}\, M_\odot$)~\cite{Carr:2009jm, Katz:2018zrn} and the black dashed lines correspond to PBHs between 1 and 100 solar masses,  for $\alpha_c=1$. For the other values of $\alpha_c$ the band moves slightly to the left.}
    \label{fig:Pheno}
\end{figure}

We now proceed to discuss the phenomenological impact of our estimates of the spectrum of stochastic GWs and of the abundance of PBHs from the DW network. We base our results on the numerical output described in Sec.~\ref{sec:numerics} and on the analytical understanding of the evolution of the false vacuum fraction described in Sec.~\ref{sec:analytics}. 
We start by estimating the minimal PBH abundance, formed from  Hubble sized PBHs pockets that reach  $\alpha_\text{loc}=1$, using  eq.~\eqref{Sasy}. The total abundance may be expressed as a function of the energy fraction of the network at the time of GW emission, $\alpha_\text{gw}$, and the background temperature at that time, $T_\text{gw}$. The mass of these black holes is set roughly by the total energy in the Hubble volume and more precisely by eq.~\eqref{E}. The abundance at a given mass is experimentally constrained by a wide variety of probes. In Fig.~\ref{fig:Pheno} we translate the bounds on the PBH abundance from  \cite{Green:2020jor} and \cite{Carr:2021bzv} into bounds on the $\alpha_\text{gw}-T_\text{gw}$ plane (thick blue curve). We relate $\alpha_\text{loc}$ with $\alpha_\text{gw}$ using eq.~\eqref{fullalpha}, which is a minor refinement of eq.~\eqref{alphaloc}. Constraint on $\alpha_\text{gw}$ are obviously stronger for larger $T_\text{gw}$, as PBHs redshift as matter for a longer time.  As an interesting example of the typical mass of these PBHs, we show in green the boundaries of the asteroid mass range  $10^{-16}M_\odot \lesssim M_\text{PBH}\lesssim 10^{-11}M_\odot$ where the PBHs can account for the whole of the dark matter \cite{Carr:2020gox}.

The remaining blue lines in Fig. \ref{fig:Pheno} refer to tentative estimates of the abundance of PBHs, 
by assuming some benchmark values, $\alpha_c=0.3$ and $\alpha_c=0.1$, at Hubble crossing, which indicates how much the structure has to further shrink (without dissipation) to enter its Schwarzschild radius.

In the same plot, we also show the range of parameters where the GW signal from the DW network could be observed by different GW detectors, from SKA~\cite{Janssen:2014dka} at the lowest frequencies to LIGO-Virgo-KAGRA (LVK)~\cite{KAGRA:2013rdx}, ET and CE at the highest ones. We have fixed the frequency of the GW spectrum at the peak to be dictated by the Hubble radius at $T_\text{gw}$, and amplitude given by eq.~\eqref{eq: GW abundance}, with efficiency $\epsilon=0.6$,  as obtained from the numerical results, and assumed the spectrum to decrease as $1/\omega$ for frequencies $\omega$ larger than the peak and $\omega^3$ for smaller frequencies. This behavior corresponds to that observed in the scaling regime~\cite{Hiramatsu:2013qaa}, and has been roughly confirmed recently also during the annihilation phase by~\cite{Kitajima:2023cek}. The GW spectra computed in our simulations roughly agree with those works, although a dedicated study is necessary to firmly establish the high frequency slope.

We then plotted the regions of parameters where the spectrum overlaps with the power law sensitivity curves derived in \cite{Schmitz:2020syl}. In the same figure, we also show the current bounds obtained in~\cite{KAGRA:2021kbb} with LIGO-Virgo O3 data (LV), which already indirectly constraints the maximal PBH abundance for $\alpha_c=1$, and the region of parameter space which could provide an interpretation of the PTA signal (red contours)~\cite{NANOGrav:2023hvm}.

Let us focus on the PTA region first, i.e. at $T_{\text{gw}} \approx 1-10$ GeV. We find the DW interpretation of the PTA signal to be overall compatible with constraints on PBHs for collapse thresholds $\alpha_c\gtrsim 0.1$. Interestingly, if $\alpha_c\lesssim 1$, a fraction of PBH dark matter from DW collapse is expected,  which is compatible with what is inferred from the BH merger rate measured by LIGO/Virgo. Significant tension between PTA observations and astrophysical bounds on PBHs would instead arise only if the collapse threshold were even smaller, i.e. $\alpha_c\lesssim 0.1$. The likelihood of such low thresholds remains to be assessed by future work. Our results differ drastically both from~\cite{Gouttenoire:2023gbn} (for example regarding the time dependence of the DW decay) and from the earlier contradictory in the first version of~\cite{Gouttenoire:2023ftk}.

As mentioned above, a particularly interesting region of parameter space is the asteroid mass range, for  $10^6  \text{ GeV} \lesssim T_{\text{gw}} \lesssim 10^9 \text{ GeV}$, where the totality of dark matter could be explained by PBHs from the network. This mass range is typically very hard to probe, given the particle-like size of their Schwarzschild radius. Crucially however, if PBHs originate from the DW network, a complementary GW signature of their existence is expected, that can be probed partially by LVK at design sensitivity and fully by ET and CE. 

Beside these two interesting regions, the plot shows that in a large range of parameters, the annihilation of the network can be ``heard'' by different GW observatories, and that a non-negligible abundance of PBHs might also be expected if $\alpha_\text{gw}> {\cal O}(10^{-2})$.

The interplay of GW and PBHs signatures described in this work is similar to the more studied scenario of PBH formation from the collapse of large adiabatic perturbations from inflation.  However, GWs from density perturbations as an interpretation of PTA data have been shown to be in tension with constraints from PBH overproduction~\cite{Dandoy:2023jot, Franciolini:2023pbf}. In contrast, the mechanism presented here remains viable according to our current understanding. Additionally, asteroid-mass PBH DM from inflationary perturbations has been argued to give a GW signal peaked in the LISA frequency band~\cite{Bartolo:2018evs}, whereas ground-based interferometers (LVK, ET, CE) are best-suited to indirectly probe PBH DM from DW collapse.

%%%%%%%%%%%%%%%%%%%%%%%%%%%%%%%%%%%%%%%%%%%%%%%%%%%%%%%%%%%%%%%%%%%%%%%

%\acknowledgments
\begin{acknowledgments}
We thank Daniel Figueroa and Francisco Torrent\'i for help with {\tt Cosmolattice}. We thank Juan Sebast\'ian Valbuena-Berm\'udez for spotting some typos in the first version of this paper.
The work of F.R.~is partly supported by the grant RYC2021-031105-I from the Ministerio de Ciencia e Innovaci\'on (Spain). F.R. thanks the MITP in Mainz (DE) for kind hospitality during completion of this work. R.Z.F. acknowledges the financial support provided by FCT – Fundação para a Ciência e Tecnologia, I.P., reference 2022.03283.CEECIND as well as the FCT projects CERN/FIS-PAR/0027/2021, UIDP/04564/2020 and UIDB/04564/2020. 
The work of A.N. is supported by the grants PID2019-108122GB-C32 from the Spanish Ministry of Science and Innovation, Unit of Excellence Mar\'ia de Maeztu 2020-2023 of ICCUB (CEX2019-000918-M) and AGAUR2017-SGR-754. A.N. is grateful to the Physics Department of the University of Florence for the hospitality during the course of this work. OP acknowledge the support from the Departament de Recerca i Universitats from Generalitat de Catalunya to the Grup de Recerca ‘Grup de Física Teòrica UAB/IFAE’ (Codi: 2021 SGR 00649) and the Spanish Ministry of Science and Innovation (PID2020-115845GB-I00/AEI/10.13039/501100011033). IFAE is partially funded by the CERCA program of the Generalitat de Catalunya.
\end{acknowledgments}

\appendix 
\section*{Appendix}

\section{Numerical strategy}
\label{app:numerics}

The comoving lattice spacing is $\Delta x = L/N$, while
the physical lattice spacing is $\Delta x_\text{phys}=a(\eta)\Delta x$. Units are chosen such that $v=1$, and  $m=\sqrt{2\lambda} v=1$, which corresponds to fixing $\lambda=1/2$. The initial value of $a(\eta)$ is fixed to 1. In radiation domination, $a\propto 1+H_i\eta$, $H= H_i(1+H_i \eta)^{-2}$ and $ \mathcal{H}\equiv a H\propto a^{-1}\propto \eta^{-1}$.
 
For Domain walls the following numerical conditions have to be satisfied:
\begin{itemize}
\item{The physical DW width $\delta_w$ should remain larger than the physical lattice spacing until the final time, i.e.
\begin{equation}
\delta_w\sim m^{-1}\gg \Delta x_\text{phys} (\eta_f) \Rightarrow H_i\eta_f \ll \frac{N}{L} \, .
\end{equation}}
\item{There should be $c\gg 1$ Hubble patches at the final time of the simulation in the physical box, 
\begin{equation}
\left(\frac{\Delta x_\text{phys}(\eta_f) N}{H^{-1}(\eta_f)}\right)^3\sim c \Rightarrow L \gtrsim c^{1/3} \eta_f \, .
\end{equation}}

\end{itemize}
In the extreme allowed case, the width of the domain wall is of the order of the lattice spacing at the end of the simulation. Then putting together the conditions above we get
\begin{equation}
N\gtrsim c^{1/3} \eta_f^2.
  \end{equation}
For instance, if we want to get to $\eta_f\gtrsim 25$ with $c\sim 5$ Hubble patches at the end of the simulation, then we need $N\gtrsim 1100$. If we want to get further, like $\eta_f=30$, then we need $N\gtrsim 1500$. The longest simulation time is obtained if both conditions are violated at the same time, which corresponds to setting $L=\sqrt{N/H_i}=\eta_f$.

\section{FV pockets in the Nambu Goto approximation}
\label{section:NG}

The thin wall approximation provides additional insight on the DW network behaviour, especially in the annihilation phase where the remains of the network is a collection of separate FV pockets. This is a good approximation when the DW worldsheet curvature radius, $R$, is large compared to its width, which is set by the inverse scalar mass. (This is satisfied  in most of the network during scaling, except for a small fraction of the total volume where collisions, interconnections or pinch-off events occur). It is possible to include the gravitational effect from the DWs themselves (see {\em e.g.} \cite{Garriga:2015fdk,Deng:2016vzb,Deng:2017uwc,Deng:2020mds}) but we shall ignore this here.

In this approximation, the evolution of a FV pocket follows from the `equation of motion' for the DW at its boundary, which reduces to the Nambu-Goto (NG) equation
$$\sigma \, K = \Delta V~.$$ 
Here $K$ is the extrinsic curvature of the DW worldsheet and we included a pressure term given by the bias $\Delta V$, see e.g.~\cite{Guven:1993ew}. It is not easy to solve this equation in general. However, the equation simplifies for walls with higher symmetry. We are interested in the motion of a DW network where inevitably the DW shapes are random. However, once the network annihilation starts, the DW motion, in a way, simplifies. Indeed, the DWs are simply the boundaries of FV pockets, which shrink quite quickly and quite independently of their initial shape.

This can be illustrated by comparing 3 extreme cases that can be easily computed, where the shape of the DW is: i) spherical, ii) cylindrical and iii) planar. The NG equation then reduces to 
\begin{equation}\label{NG}
R''+ \left(\frac{n}{R}- 3 R' \frac{a'}{a}\right) \gamma^{-2} 
+a\,\frac{\Delta V}{\sigma} 
\gamma^{-3}=0 \, ,
\end{equation}
with $n=2,1$ for spherical or cylindrical DW of comoving radius $R$ respectively. The case $n=0$ corresponds to a planar wall placed at, say,  $z=R(\eta)$. Primes denote derivatives w.r.t. conformal time, and 
$$
\gamma\equiv 1/\sqrt{R'^2-1}~.
$$

It is straightforward to integrate this equation and thus follow the evolution of a structure of certain initial comoving radius $R_0$. Some representative examples are shown in Fig.~\ref{fig:traj}. It is clear from the figure that the structures reach arbitrarily small size after a finite (conformal) time $\Delta \eta$, that is the time lapse until $R$ approaches $0$. Of course eventually a small enough structure transfers its energy into scalar waves (which are not captured in the NG approximation). Note that if one prefers to define the collapse time as when $R(\eta)$ reaches a small radius $r_c$, the result would be the same so long as $r_c \ll R_0$.

\begin{figure}[h]
\includegraphics[width=.48\textwidth]{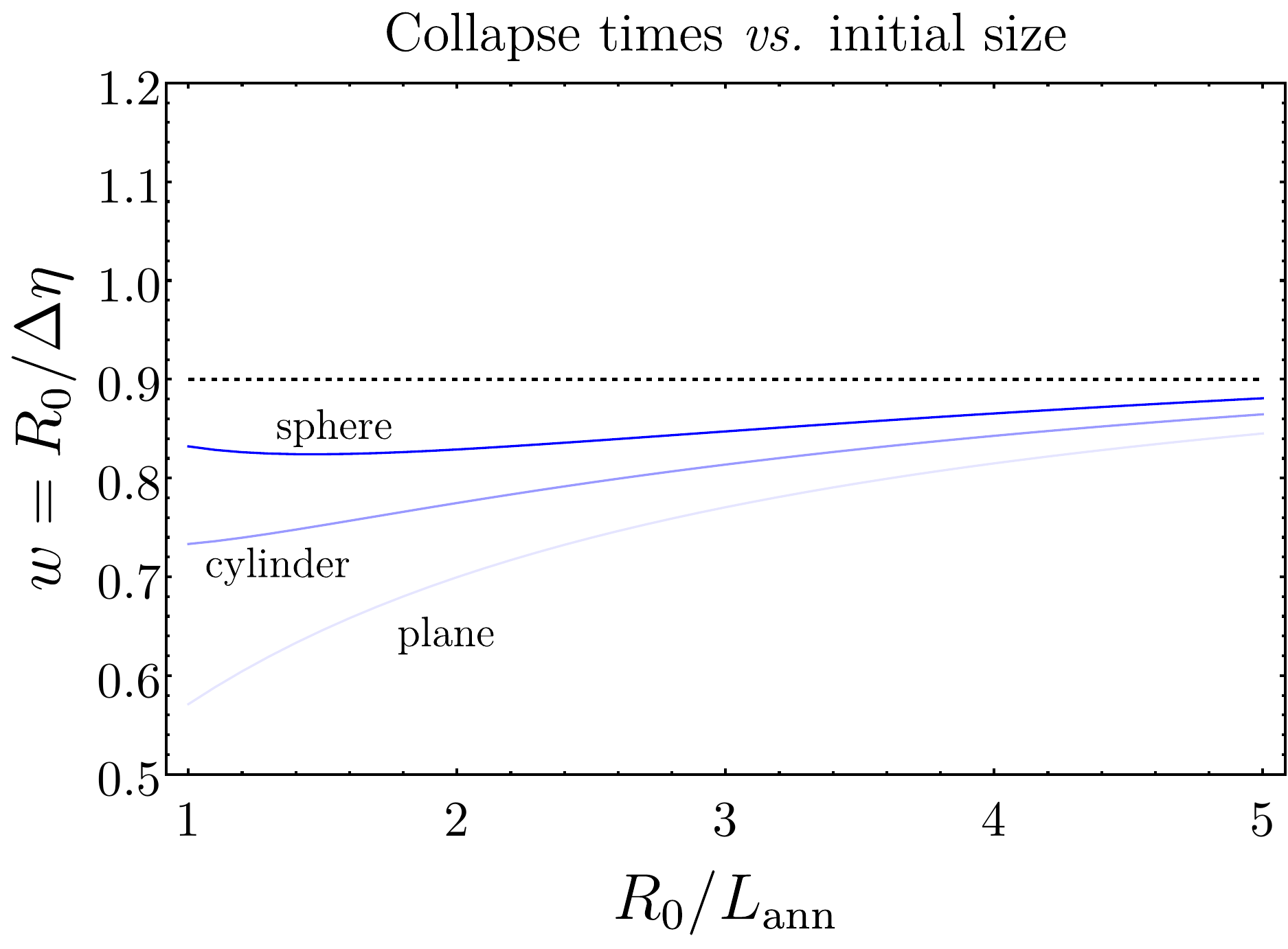}    
\caption{
Ratio of the initial radius $R_0$ and the (conformal) time to reach $R(\eta)=0$, $\Delta \eta$,  of super-Hubble FV pockets. It depends mildly on the shape and it asymptotes to a constant value (dotted black curve).
} 
\label{fig:collTime}
\end{figure}

The trajectories shown in Figs.~\ref{fig:traj} are readily understood: closed DWs shrink under both the effect of the tension and the pressure $\Delta V$, reaching relativistic speeds quite quickly.

Interestingly, the collapse time $\Delta \eta$ depends mostly on the initial size. As shown in Fig.~\ref{fig:collTime}, for large enough initial FV pockets, the collapse time approaches $C \,R_0$, with a constant $C$ in the range $1.15-1.2$, quite independently of the pocket shape. 

Of course, in a network the DW shapes are not symmetric. However, Fig.~\ref{fig:collTime} signals a quite clear time scale in the network decay: the first stage of annihilation, (that is still far from the FV pocket gas picture) takes about one Hubble time.

One expects a `burst' of GWs from the collapsing FV regions as they shrink because they are significantly non-spherical. This GW production time $\eta_\text{gw}$ is then expected to be near the collapse time because it is dominated by the most numerous pockets, with sizes of order $\eta_{\Delta V}$. In summary, the fact that Hubble-sized structures have to reach small sizes naturally leads to the expectation
\begin{equation}\label{delayNG}
\eta_\text{gw} \simeq \eta_{\Delta V} + (\Delta \eta)|_{R_0\sim \eta_{\Delta\text{V}}} \sim 2 \eta_{\Delta V}~.    
\end{equation}

It is also easy to keep track of the energy of a given FV pocket, and how it evolves in time. Focusing on spherical symmetry for simplicity, the energy is
\begin{equation}\label{E}
E=\frac{4\pi}{3} \Delta V R^3(\eta) a^3(\eta) + 4\pi \sigma R^2(\eta) a^2(\eta) \gamma(\eta)~,    
\end{equation}
and because of the expansion it is not conserved. For super-Hubble pockets, first $E$ grows (because both the FV region and the DW gain volume/area by the expansion). Only when they enter the Hubble radius the energy stabilizes.  

We can keep track of the energy carried by the FV pocket that enters the Hubble radius at each time, by evaluating eq.~\eqref{E}. 
The $\gamma$ factor grows in time, but not enough to compare to the volume contribution. Thus, after about one Hubble time, $E$ scales like the physical Hubble volume, $\sim\eta^6$. Solving numerically the NG equation \eqref{NG} we find that the expression $E/E_0 \simeq (\tau^6+\tau^5+2\tau^4)/4$, with $\tau=\eta/\eta_\DV$ and $E_0$ the initial energy of the pocket, provides a good fit (within $5\%$) of the actual time dependence (the $\tau^5$ term can be identified as the DW gamma factor). This leads to the following improvement of eq.~\eqref{alphaloc}
\begin{equation}
\alpha_\text{loc} = 1.5 \, \alpha_\text{gw} \frac{\tau^4+\tau^3+2\tau^2}{\tau_\text{gw}^4+\tau_\text{gw}^3+2\tau_\text{gw}^2}    \, ,
\label{fullalpha}
\end{equation}
where the factor of 1.5 comes from rewriting  $\alpha_\text{loc}(\eta_\text{gw}) =\Delta V/(3 H(\eta_\text{gw})^2 M_p^2)$ in terms of the fraction of the energy density in the DW network at the same time $\alpha_{\text{dw}}= \rho_{\text{dw}}/\rho_c$, with $\rho_{\text{dw}}(\eta_\text{gw})=2.6 \, \sigma H(\eta_\text{gw})$ and $\tau_\text{gw}\simeq 2$. This is smaller than eq.~\eqref{alphaloc} for $\tau>\tau_\text{gw}$, so eq.~\eqref{alphaloc} provides a conservative bound.

%%%%%%%%%%%%%%%%%%%%%%%%%%%%%%%%%%%%%%%%%%

% \bibliographystyle{JHEP}
% \bibliography{biblio.bib}
\bibliography{paper.bib}

\end{document}